\documentclass{ametsoc_nolinenumbers}

\journal{jamc}

\usepackage[german,english]{babel}
\usepackage{xcolor,colortbl}
\usepackage{enumerate}
\usepackage{booktabs}
	
\usepackage[latin1]{inputenc}
\usepackage[T1]{fontenc} 
\usepackage{bbm}

\newcommand{\hdcp}{HD(CP)$^2$ }
\newcommand{\cb}{\it}
\newcommand{\cy}{\bf}
\nolinenumbers
\bibpunct{(}{)}{;}{a}{}{,}

\title{Using the SAL technique for spatial verification of cloud processes:\\A sensitivity analysis}

    \authors{\nolinenumbers Michael Weniger\correspondingauthor{\nolinenumbers Michael Weniger, Meteorological Institute, University of Bonn, 
				Auf dem H\"ugel 20, 53121 Bonn, Germany.}
 and Petra Friederichs\thanks{Current affiliation: Meteorological Institute, University of Bonn,  Germany.}}
    \affiliation{\nolinenumbers Meteorological Institute, University of Bonn,  Germany}

\email{mweniger@uni-bonn.de}

\abstract{\small\nolinenumbers The feature based spatial verification method SAL is applied to cloud data, i.e. two-dimensional spatial fields of total cloud cover and spectral radiance. Model output is obtained from the COSMO-DE forward operator SynSat and compared to SEVIRI satellite data. The aim of this study is twofold. First, to assess the applicability of SAL to this kind of data, and second, to analyze the role of external object identification algorithms (OIA) and the effects of observational uncertainties on the resulting scores. \\
As a feature based method, SAL requires external OIA. A comparison of three different algorithms shows that the threshold level, which is a fundamental part of all studied algorithms, induces high sensitivity and unstable behavior of object dependent SAL scores (i.e.\ even very small changes in parameter values can lead to large changes in the resulting scores). An in-depth statistical analysis reveals significant effects on distributional quantities commonly used in the interpretation of SAL, e.g.\ median and interquartile distance. Two sensitivity indicators based on the univariate cumulative distribution functions are derived.  They allow to asses the sensitivity of the SAL scores to threshold level changes without computationally expensive iterative calculations of SAL for various thresholds. The mathematical structure of these indicators connects the sensitivity of the SAL scores to parameter changes with the effect of observational uncertainties.\\
Finally, the discriminating power of SAL is studied. It is shown, that -- for large-scale cloud data -- changes in the parameters may have larger effects on the object dependent SAL scores (i.e.\ the $S$ and $L2$ scores) than a complete loss of temporal collocation.}

\begin{document}

\maketitle

%

\section{Introduction}
\label{sec:introduction}

Verification of numerical model output is essential in the development of successful models for numerical weather prediction. Due to an increase in model resolution new techniques for the evaluation of spatial fields have emerged during the last decade (see for instance \citet{casati2008forecast,gilleland2009,ebert2008}). Feature based methods are an important part of this toolkit. These methods use score functions that are defined on objects, and not on the spatial field itself (i.e.\ on a subset of the spatial data usually identified by some external algorithm). Most feature based methods have been designed with a specific field of application in mind. Verification of precipitation fields is the most prominent application, and various methods have been developed for this kind of spatial data, e.g.\ Contiguous Rain Area \citep{ebert2000verification,ebert2009toward}, Method for Object-based Diagnostic Evaluation \citep{davis2006object,davis2006object2} and SAL \citep{wernli2008sal}. 

SAL stands for its three score components: (S)tructure, (A)mplitude and (L)ocation. It was developed to measure the quality of a forecast using three distinct scores, which have direct physical interpretations to allow for conclusions on potential sources of model errors. It does not require matching individual objects in observations and forecasts, but compares the statistical characteristics of those fields. The resulting scores are close to a subjective visual assessment of the accuracy of the forecast for precipitation data. SAL was originally developed for the verification of precipitation fields in a defined area (e.g. river catchments) and today it is widely used in the evaluation of quantitative precipitation forecasts \citep[e.g.][]{zacharov2013evaluation,leoncini2013ensemble,zimmer2011classification}. Recently, efforts to employ SAL to different kinds of data have been made: \citet{shi2014improved} used SAL for the evaluation of a soil moisture model and \citet{crocker2013exploratory} applied SAL on binary cloud masks.

The aim of this work is twofold: first, to assesses the benefits and drawbacks of SAL applied on cloud data. And second, to systematically study the role of the object identification algorithms (OIA) and their parameters. Spatial fields that describe CP processes such as  total cloud cover or spectral radiance may contain large scale structures. A focus of this study is thus to investigate how well SAL is able to deal with large features, and to quantitatively analyze the effect of different OIA parameter settings. \citet{wernli2008sal} investigated the so-called ``camel cases'' on a qualitative level and showed that even small changes in the threshold level of the OIA can lead to very different SAL scores. We follow this line of thought and conduct an extensive statistical analysis of large sets of spatial cloud data to quantify the sensitivity of SAL towards three parameters: threshold ration, smoothing radius and minimal object size. Since substantially different threshold levels correspond to different physical situations, we expect the resulting scores to be different as well. This is true not only for SAL but virtually any threshold based verification method, such as the Fractions Skill Score \citep{roberts2008scale} or the Intensity-Scale Skill Score \citep{casati2004wavelet,casati2010new}. The interesting question is how the scores react to very small changes in parameter values, i.e. whether the verification score is numerically stable with respect to its OIA parameters.

This stability, i.e. the effect of small perturbations in parameter values, is closely linked to the effect of observational uncertainties, i.e. small perturbations in the data itself. Observational uncertainties are generally ignored in spatial verification methods (\cite{ebert2013progress} and references therein), which might be justified if observational errors are small compared to model errors. However, this assumption is not true for remotely sensed estimates of variables (e.g. estimates derived from radar or satellite observations), particularly those related to cloud and precipitation (CP) processes. While the instrument errors of direct satellite measurements such as spectral radiance or brightness temperature are small, this is not true for derived quantities such as cloud fraction or cloud masks \citep{zinner2005remote,crocker2013exploratory}. Additional uncertainties enter the verification process in form of spatial interpolation due to the discrepancy between the model grid and the, usually irregular, observational grid. The evaluation of CP processes in high-resolution model simulations strongly relies on this kind of remotely sensed observations \citep{evaristo2014,steinke2015,hammann2015,nam2014,eggert2015temporal}. 
Therefore, it is important to understand the behavior of SAL with respect to observational uncertainties.

%

The article is structured as follows. We first provide the mathematical definitions of SAL in Section \ref{sec:definition-of-sal}. Three different OIA and conceptional scenarios, which allow us to identify focal points for the analysis of SAL's parameter sensitivity, are discussed in Section \ref{sec:oia}. These points are explored with exemplary cases and an in-depth statistical analysis using spatial data of total cloud cover and spectral radiance in Section \ref{sec:analysis-of-parameter-sensitivity}. The threshold parameter is of particular importance, since it is the basis of all three OIA, is closely connected to observational uncertainties and impacts not only SAL but other threshold based verification techniques. A-priori and a-posteriori indicators, which provide a computationally effective way to asses SAL's sensitivity to varying thresholds, are discussed in Section \ref{sec:indicators}. The insights gained from the mathematical formulation of these indicators are used to establish the link between parameter sensitivity and observational uncertainties. Section \ref{sec:randomly-permuted-forecasts} investigates the ability of the object dependent SAL scores to  distinguish between two different sets of cloud data.

\section{Definition of SAL}
\label{sec:definition-of-sal}
Let us consider a two-dimensional domain $\mathcal{D}\subset{\mathbb{R}^2}$ composed of $N\in\mathbb{N}$ grid points with a maximal diameter
\begin{align*}
	d := \sup_{(x,y)\in \mathcal{D}}|x-y| > 0.
\end{align*} 
We now want to evaluate one set of spatial data $R_1$ on the domain $\mathcal{D}$ with respect to a second set of data $R_2$. 
To this end, for each field $R_i$, we define $n_i\in \mathbb{N}$ objects $O_{i,k}\subset\mathcal{D}$ with $k\in\{1,\ldots,n_i\}$, $i\in\{ 1, 2 \}$ using some OIA. The OIAs are discussed later in Section \ref{sec:oia}. 
Based on the defined objects $O_{i,k}$ the three components of SAL are defined as follows:
\begin{itemize}
  \item (A)mplitude
    \begin{align*}
	A &= \frac{\langle R_1\rangle_\mathcal{D}- \langle R_2\rangle_\mathcal{D}}{0.5\left( \langle R_1\rangle_\mathcal{D}+ \langle R_2\rangle_\mathcal{D} \right)} \in [-2,2],
    \end{align*}
    where $\langle . \rangle_\mathcal{D}$ denotes the average over the domain $\mathcal{D}$. A perfect A-score of $A=0$ indicates that $R_1$ is  unbiased with respect to $R_2$. In the case of $A=1$, the spatial data $R_1$ is overestimated by a factor of 3, wheres $A=-1$ means that $R_1$ is underestimated by a factor of 3. 
  \item (L)ocation\\
    Let us denote the center of total mass for the field $R_i$ by $x_i\in\mathbb{R}^2$, for $i\in\{ 1, 2 \}$, the center of mass for each object $O_{i,k}$ by $x_{i,k}$ and the mass of each object by $M_{i,k}\in\mathbb{R}$, for $k\in\{1,\ldots,n_i\}$, $i\in\{ 1, 2 \}$. The L-score is defined as
    \begin{align*}
	L1 &= \frac{| x_1-x_2 |}{d} \in [0,1] \\
	r_i &= \frac{\sum_{k=1}^{n_i} M_{i,k}|x_i-x_{i,k}|}{\sum_{k=1}^{n_i} M_{i,k}}\quad\text{\small scattering of objects} \\
	L2 &= 2\frac{| r_1-r_2 |}{d} \in [0,1] \\
	L &= L1 + L2 \in [0,2] .
    \end{align*}
    The first part of the L-score, $L1$, describes the relative distance between the centers of total mass $x_1$ and $x_2$. The second part, $L2$, is a measure for the scattering of the identified objects. Since both L-scores are fully defined by the centers of total mass and centers of object's mass, $L$ is rotation invariant, i.e. rotating the whole field or an object around its center of mass does not change the $L$ score. For $L=0$ we have a perfect location match of all centers of mass.
  \item (S)tructure
    \begin{align*}
	V_{i,k} &= \frac{M_{i,k}}{\max_{x\in O_{i,k}}R_i(x)}\qquad\ \text{\small scaled mass of object }O_{i,n} \\
	V_i &= \frac{\sum_{k=1}^{n_i} M_{i,k} V_{i,k}}{\sum_{k=1}^{n_i} M_{i,k}}\quad \text{\small scaled, weighted total mass} \\
	S &= \frac{V_1-V_2}{0.5\left( V_1 + V_2 \right)} \in [-2,2] 
    \end{align*}
   $V_{i,k}$ is the mass of the object $O_{i,k}$ after it has been rescaled to maximal height of $1$. The scaled total mass $V_i$ is the weighted and normalized sum over the rescaled masses. The intent of the rescaling is to remove, or at least dampen, the influence of total mass and concentrate on the structure of the objects. An S-score of $S=0$ is obtained for a perfect match for the structures of all objects in both data sets. If $S<0$  then the objects of $R_1$ are too peaked compared to those of $R_2$, whereas for a positive S-score, $S>0$, implies that they are too flat. 
\end{itemize}
For visualizations of the SAL properties the reader is referred to \citep{wernli2008sal}.

\section{Object Identification Algorithms}
\label{sec:oia}
One central component of SAL is the identification of objects. While $A$ and $L1$ scores are independent of the object identification, since they are directly defined on the $R_i$ fields, $L2$ and $S$ are defined on the sets of objects $O_{i,k}$. 
As there exists a variety of OIA, which in turn require the specification of parameters, it is imperative to understand the sensitivity of the SAL scores to the choice of the OIA. We start this study with the discussion of some conceptual cases. These theoretical consideration show potential issues, which may lead to very unstable responses of SAL with respect to small changes in parameter values. We will study the parameter sensitivity of three different OIA, which are described in the following section, by looking at the behavior of the object dependent scores $S$ and $L2$. 

In order to identify cohesive objects in spatial data, OIA typically use a threshold level and define a continuous set of points of threshold exceedances as one object. In order to filter small scale noise many methods apply a smoothing filter prior to object identification, or ignore objects smaller than a predefined number of points. From the multitude of existing OIA we apply three methods implemented in the R package \textit{SpatialVx} \citep{spatialvx}.
 
The OIA \textit{threshfac} is the algorithm originally used with SAL \citep{wernli2009spatial}. It defines an object as a cohesive set of threshold exceedances. The threshold is defined as $R^*_i = f\ \cdot \ R^{95}_i$, where $R^{95}_i$ is the $95\%$-quantile of the field $R_i$, $i\in\{ 1, 2 \}$ and $f > 0$ a threshold ratio with a default value of $f=1/15$. This simplistic approach has the advantage that its only parameter, the threshold level, has a direct physical interpretation. The lack of smoothing or filtering makes this method susceptible to the effect of small scale noise, which might lead to a unrepresentative dominance of very small, scattered objects. This issue is addressed by the following OIA.

The convolution threshold algorithm \textit{convthresh} \citep{davis2006object,davis2006object2} identifies objects in two steps. First, the data fields are convolved with a smoothing process, i.e. the value at each grid point is replaced by the mean value over a disc with a radius given by the parameter \textit{smoothpar}. Second, the convolved data is thresholded yielding a binary mask, which in turn is applied to the original fields. The resulting objects thus have the original values at each grid point but smoothed boundaries. The advantage is that the borders of the objects are smooth similar to those a human would draw manually. The method filters out small scale noise (i.e. small scattered objects), which is either isolated or located at the borders of large objects. The drawback is the introduction of an additional parameter, the smoothing radius, which has no direct physical interpretation. Therefore it is not obvious how to choose this parameter for a given set of data.

The algorithm \textit{threshsizer} \citep{nachamkin2009application} defines objects as cohesive threshold exceedances, where objects consisting of less than \textit{NContig} grid points are omitted. This method is used to filter out small isolated objects. The interpretation of the \textit{NContig} parameter is more straightforward than the smoothing radius in the previous OIA, but has no effect on the shape of large objects. Hence, it is easier to foresee the consequences of a particular choice of parameter value, but the objects do not look as \textit{natural} as the ones provided by \textit{convthresh}.\\

\begin{figure}
\center
\includegraphics[width=1.0\textwidth]{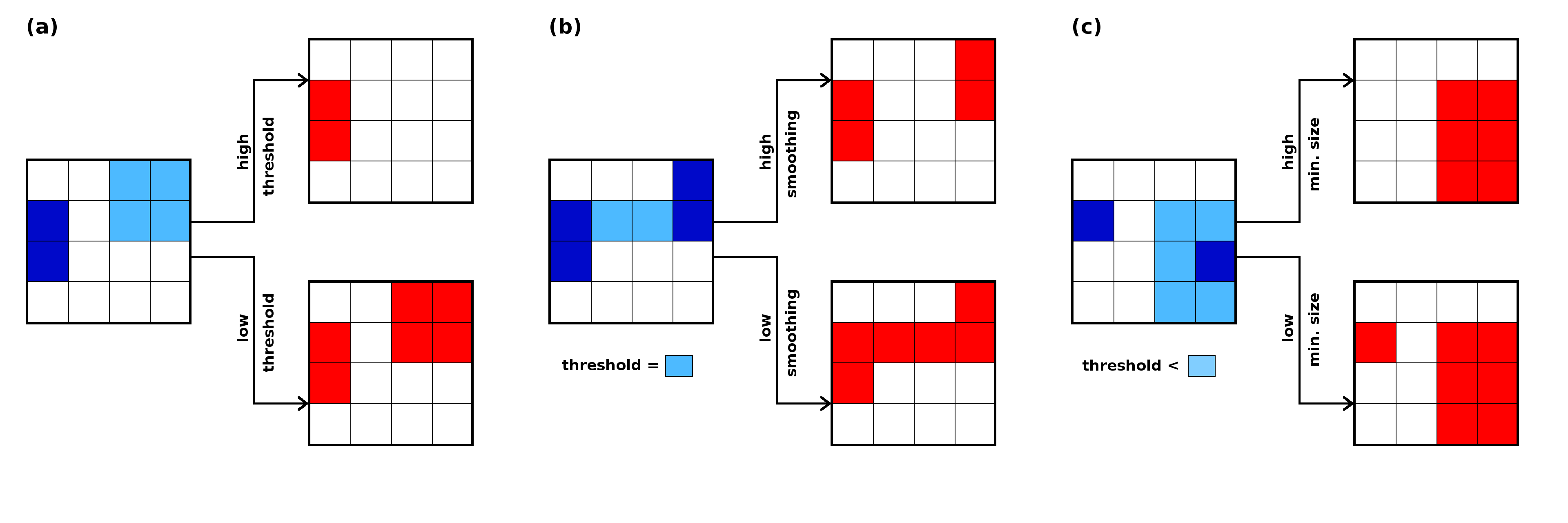}
  \caption{Conceptual OIA cases. The left part of each panel shows the data of a domain with $4\times 4$ grid points. Dark blue squares indicate points with high intensity, while light blue coloring denotes an intensity value near the threshold level. The resulting object masks are plotted in red. Panel (a) shows varying threshold levels that cause a large object to vanish, (b) demonstrates object decomposition due to varying smoothing radii, and (c) shows varying minimal object sizes that cause a small object to vanish.}
  \label{fig:thoughtexp}
\end{figure}

Let us now consider conceptual cases on a $4\times 4$ grid with values of different intensity. Fig.~\ref{fig:thoughtexp} a) shows the effect of a varying threshold level. Depending on the threshold level, objects of lower intensity are either identified or ignored. The presence or absence of the lower intensity objects influences $L2$ and $S$. On the one hand, this is a deliberate effect, since different threshold levels concentrate the analysis to different physical situations. However, this effect might become problematic when the threshold is close to the lower intensity value, since an arbitrarily small change in the threshold level may cause the whole object to vanish and lead to a potentially large change in the $L2$ and $S$ scores.

The effect of the smoothing radius is illustrated in Fig.~\ref{fig:thoughtexp} b). Smoothing may cause a low intensity bridge to fall below the threshold. In our example, smoothing is achieved by averaging over a $3\times 3$ window. The resulting sets of objects differ in average spread and structure, which in turn leads to changes in the $L2$ and $S$ scores. If such a bridge is very narrow or its value is close to the threshold level, then even small changes in the strength of the smoothing may have large effects on the $L2$ and $S$ scores.

In contrast, the effect of the minimal object size parameter (i.e. the \textit{Ncontig}  of the \textit{threshsizer} algorithm) only affects small objects, regardless of the intensity of the values (Fig.~\ref{fig:thoughtexp} c)).  Since all object dependent SAL scores are weighted with the mass of the objects, the effect on the scores should be small for small changes in the parameter value. Extreme cases, where e.g.\ only two small objects are present and one of them vanishes due to a slightly raised \textit{Ncontig} parameter, may be thought of, but are very unlikely to occur for actual data with hundreds or thousands of grid points.

\begin{figure}
\center
 \includegraphics[width=1.0\textwidth]{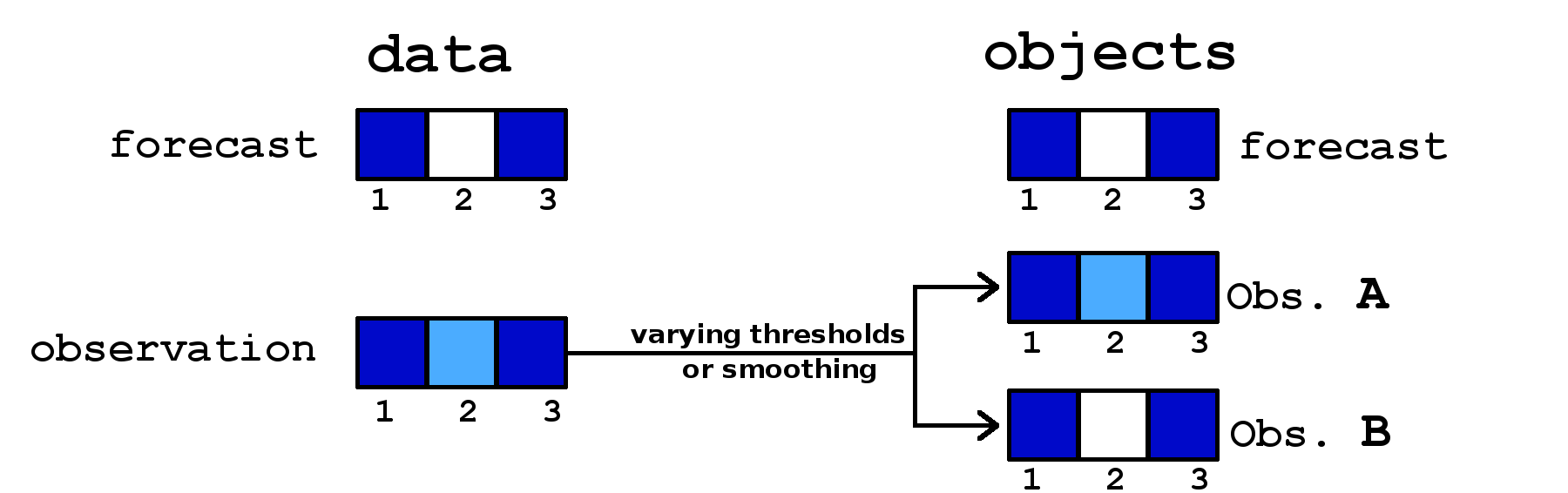}
  \caption{Conceptual case demonstrating a potential effect of small changes in parameter value. Dark blue squares indicate points with high intensity, while light blue coloring denotes an intensity value near the threshold level. \textit{Obs. A} yields the worst possible $L2=1$ score, while \textit{Obs. B} is a perfect match with $L2=0$.}
  \label{fig:sal_object_decomp}
\end{figure}

Let us consider a conceptual setting (Fig.~\ref{fig:sal_object_decomp}), which allows for an easy calculation of the $L2$ score. In this setting, the middle grid point has a value equal to the threshold. An arbitrarily small change in the threshold level yields a change in the $L2$ score from the optimal to the worst possible score. In the scenario A the $L2$ score amounts to 1, whereas in scenario B forecast and observation are identical with a perfect score of $L2=0$. This example demonstrates that there exist situations in which the $L2$ score is unstable, i.e.\ it changes from the best to the worst $L2$ value for arbitrarily small parameter changes in the OIA.

\section{Analysis of Parameter Sensitivity}
\label{sec:analysis-of-parameter-sensitivity}

To investigate the sensitivity of SAL regarding cloud processes, we use model data from the ``Synthetic Satellite Simulator''  \citep[SynSat,][]{keil2006synthetic} implemented in the operational regional weather prediction model COSMO-DE at the Deutscher Wetterdienst (DWD), which computes synthetic spectral radiances and brightness temperatures of eight MSG channels \citep{crewell2008gob}. The model has a horizontal resolution of 2.8km and covers a domain with $421 \times  461$ grid points containing Germany, Switzerland and Austria. For each day, the forecast is initiated at 00UTC and yields synthetic satellite data with a temporal resolution of 15min. As observations we use data from the SEVIRI satellite \citep{crewell2008gob,reuter2009seviri} for a domain of $302 \times  202$ grid points with a maximal horizontal resolution of $3$km. 

Nine different variables are studied: total cloud cover (TClC) and eight channels of spectral radiance. TClC is derived for the observational data as the fraction of cloudy pixels in a grid box using the NWC SAF MSG v2010 algorithm, which is based on a multi-spectral thresholding technique \citep{derrien2005msg,derrien2010improvement}. The COSMO model uses a parametrization based on relative humidity in its radiation scheme and a statistical cloud scheme \citep{sommeria1977subgrid} within the turbulence model to parametrize boundary layer clouds \citep{cosmo2013users}. We have chosen spectral radiance over brightness temperature to study parameter sensitivity, since it allows us to use SAL in its original formulation (implemented in the R package \textit{SpatialVx}). For brightness temperature the definition of threshold levels based on the $95\%$ quantile is problematic, because the minimal value of the fields is far greater than zero. Therefore, only a very small range of threshold ratios around $f=1$ would yield sensible thresholds. Thus, by using spectral radiances, we avoid additional choices how to normalize the data or change the threshold routine, and can concentrate on the effects of different OIA and their parameters. If one is primarily interested in the direct verification results (e.g. to evaluate a specific model setup) and not in a technical analysis of the verification method itself, this decision should be revisited. In this case it would be interesting to compare verification scores derived from spectral radiance and brightness temperature, which essentially describe the same physical quantity. Due to the strictly monotone increasing relation between brightness temperature and spectral radiance based on Planck's law, a one to one conversion of all data points and thresholds would not change the results of the sensitivity study. We refer to the technical reports for SEVIRI for more details \citep{eumetsat1,eumetsat2}.

\begin{figure}
\center
 \includegraphics[width=.75\textwidth]{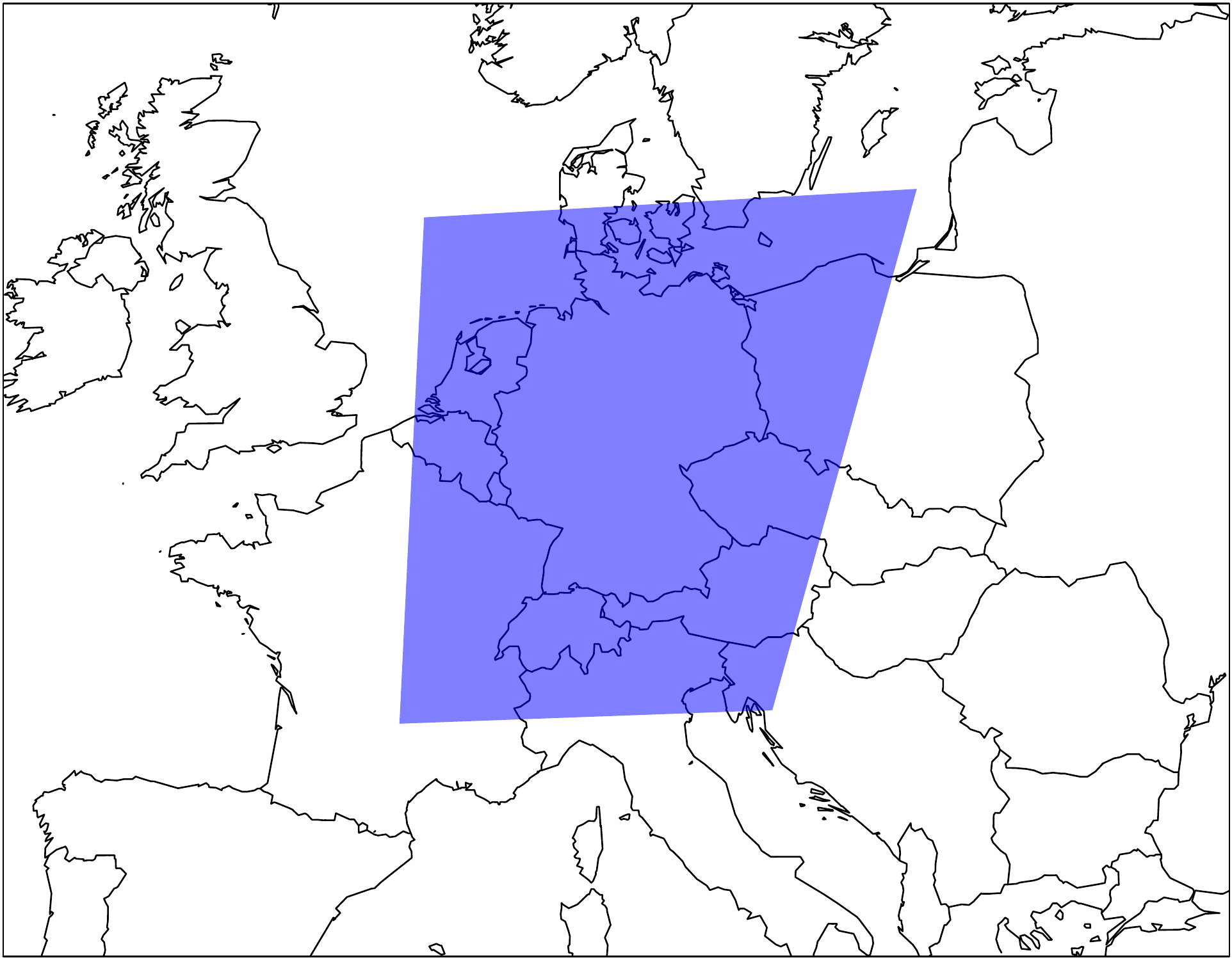}
  \caption{Map of the area shared by observational data and model output.}
  \label{fig:shared_window_map}
\end{figure}

The results shown concentrate on the spectral radiance at IR6.2, i.e.\ the water vapor band at a wavelength of 6.2$\mu$m. For each variable, we compare observed and synthetic IR6.2 radiance values every 3 hours (starting at 00UTC) between 1 January 2012 and 19 February 2012, resulting in 400 pairs of spatial fields. Note that each set of $400$ spatial fields includes forecasts of eight different lead times, which have a large impact on the verification scores. However, we are not interested in absolute SAL values, but rather in the difference of two SAL values calculated for the same fields with different OIA parameter settings. Examples in form of two case studies are stated below. The consideration of different lead times allows us to cover the whole range of small and large SAL values for the study of parameter sensitivity.
Since SAL requires all the data to be on the same grid, the model output was interpolated onto the shared area (Fig.~\ref{fig:shared_window_map}) of the coarser observational grid using a straightforward nearest-neighbor method. While the effect of different interpolation methods, e.g. bilinear, weighted, spline-based or via kriging \citep{li2008review}, may have very significant effects on verification scores, a systematic statistical analysis is out of the scope of this work. In order to focus on the study of different OIA parameter settings, we use the computationally least expensive interpolation method.

Before we explore the statistical consequences different OIA and parameter settings have on SAL, let us consider two exemplary cases, where the conceptual processes from Section \ref{sec:oia} can be observed for meteorological data. Both cases exhibit large changes in $S$ and $L2$ scores due to small changes in the parameters of the OIA. Following the line of thoughts established in Section \ref{sec:oia}, these object decomposition processes may occur for the \textit{convthresh} and \textit{threshfac} algorithm.  We distinguish between two different types of object decomposition, one where the (spatial) shape of a large object is the deciding factor, and one where the intensity structure of an object is the most important criterion. 

\begin{figure*}
\center
\includegraphics[width=\textwidth]{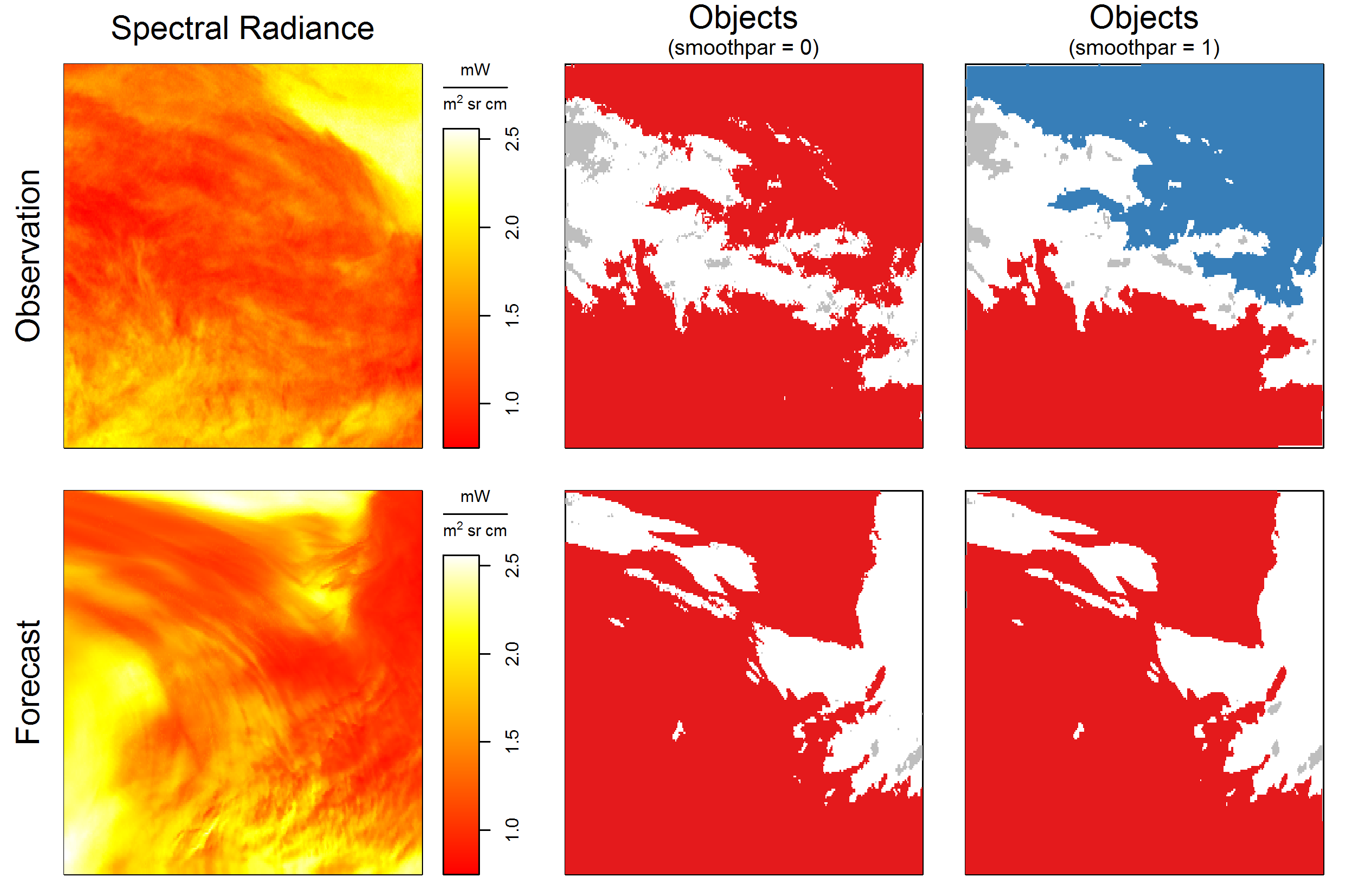}
\caption{Case study: object decomposition for the \textit{convthresh} algorithm for IR6.2 (19 January 2012, 03UTC). With a smoothing parameter of $0$ one large object is identified in the observations. Raising the smoothing radius to $1$ grid point causes the small interconnecting bridge in this object to vanish, which leads to decomposition and vastly different $S$ and $L2$ scores.}
\label{fig:case_study_rad062_convthresh}
\end{figure*}

The first type, which is shown on a conceptual level in Fig.~\ref{fig:thoughtexp}b, is responsible for most of the large deviations when using the \textit{convthresh} OIA. Fig~\ref{fig:case_study_rad062_convthresh} shows a case for IR6.2 using the \textit{convthresh} algorithm  with a smoothing radius of $0$ and $1$, respectively. Without smoothing the OIA identifies one dominating object in both observation and forecast. Although these do not match perfectly, they are very similar, which leads to small scores of $S=0.12$ and $L2=0.04$. Using the smallest possible smoothing radius of $1$ grid point causes the small interconnecting bridge in the center of the object in the observations to vanish. This leads to the decomposition of the dominant object into two large ones. Since the dominant object in the forecast is unaffected, $S$ and $L2$ scores exhibit large changes: the object in the forecast is too large, which results in a large positive structure score ($S=0.72$). The spread of objects is too small in the forecast resulting in a large $L2$ score ($L2=0.44$). In this case, the shape of the bridge is crucial, i.e. it has to be thin in order to vanish due to smoothing, while its intensity values are only of secondary importance.

\begin{figure*}
\center
\includegraphics[width=\textwidth]{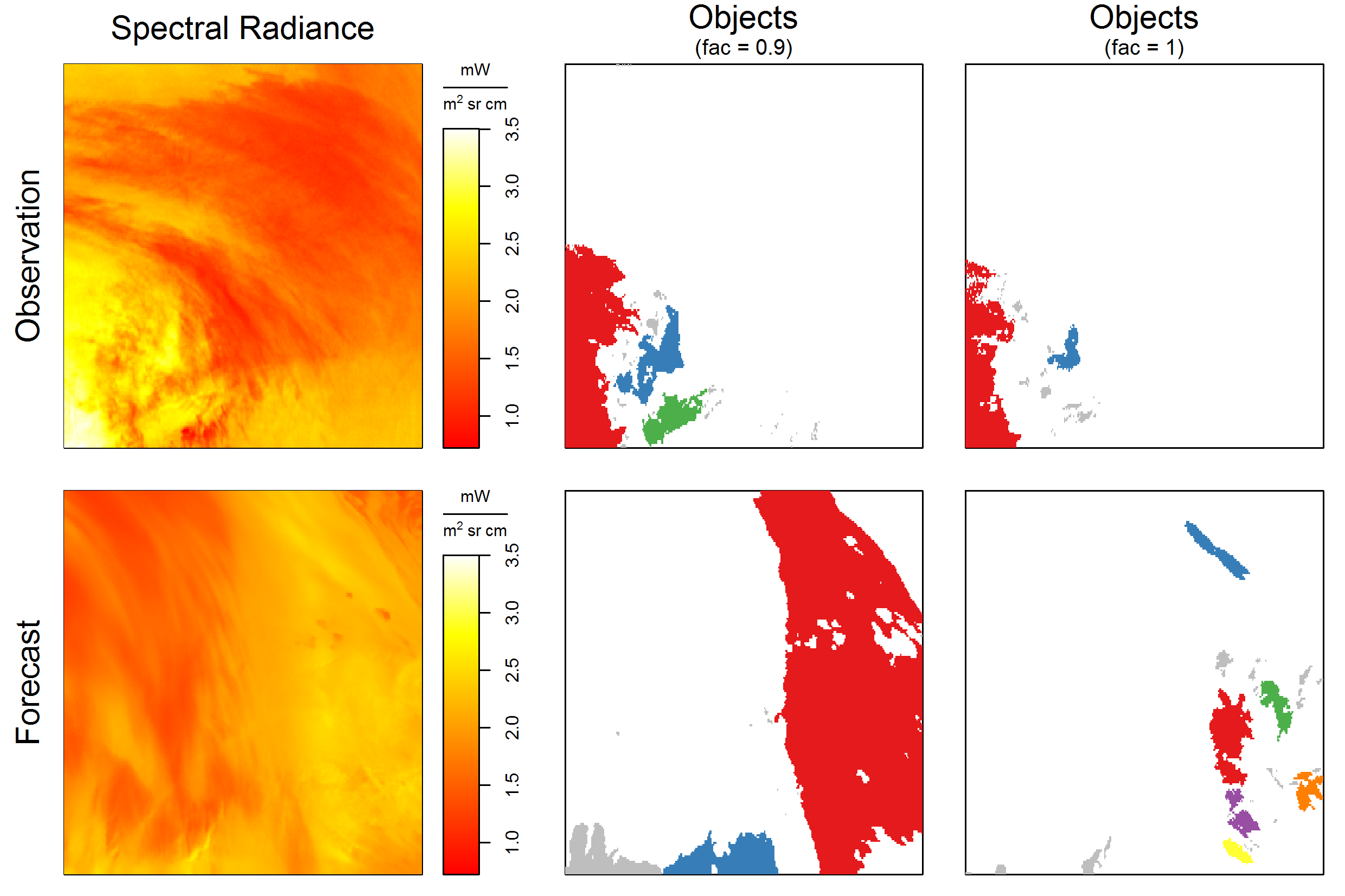}
\caption{Case study: object decomposition for the \textit{threshfac} algorithm for IR6.2 (21 January 2012, 12UTC). For a threshold ratio of $0.9$ one large object dominates the forecast. Raising the threshold ratio to $1$ causes most of the objects mass to fall below the threshold level. Effectively the object decomposes into many small ones, which leads to vastly different $S$ scores but nearly constant $L2$ scores.}
\label{fig:case_study_rad062_threshfac}
\end{figure*}

The second type of object decomposition is shown on a conceptual level in Fig.~\ref{fig:thoughtexp}a. In real meteorological data the situation is not as palpable most of the time, but the defining aspect that a large part of an objects mass falls below a varying threshold level can be clearly observed. Fig~\ref{fig:case_study_rad062_threshfac} shows a case for IR6.2 where the \textit{threshfac} algorithm is applied with threshold ratios of $0.9$ and $1$, respectively. For the lower threshold the forecast is dominated by a large elongated object. Raising the parameter causes most of this object to fall below the now higher threshold level. The remaining mass is then identified as a cluster of smaller objects. The objects in the observations become smaller but are otherwise unaffected. For the lower threshold the dominant object in the forecast is too large and thus responsible for a large positive $S$ score ($S=1.4$). The situation is reversed for the higher threshold: the structure of the clustered small objects in the forecast is too small, which leads to a large negative $S$ score ($S=-1.11$). While the effect on the $L2$ score is small in this example ($L2_{fac=0.9}=0.05$ and $L2_{fac=1}=0.18$), we have observed other cases where it exhibits large changes. The unpredictable behavior of the $L2$ score is one reason for the low correlation between absolute changes in $S$ and $L2$ scores for the \textit{threshfac} OIA, which will be discussed at the end of Section \ref{sec:analysis-of-parameter-sensitivity}.\\

The exemplary cases show that object decomposition and the resulting large changes in SAL scores can be caused by small changes in parameter values of the OIA. Let us now take closer look at the statistical effects on the distribution of SAL scores over large data sets of $N=400$ pairs of spatial fields $R_{i,j}$, where $i\in\{1,2\}$ denotes forecast and observation and $j\in\{1,\ldots,400\}$ the temporal index. We denote SAL as \textit{maximum-stable} with respect to a parameter $p$ of an OIA, if small changes in the value of this parameter ($\Delta p=p_2-p_1$) can only cause small changes in the resulting SAL scores ($\Delta SAL$), i.e.
\begin{align*}
	\max_{j\in\{1,\ldots,N\}} \left|\Delta SAL (R_{1,j},R_{2,j},p_1,p_2)\right| \leq C \left|\Delta p\right|,
\end{align*}
for a constant $C>0$. SAL is \textit{mean-stable} with respect to a parameter $p$, if there exists a constant $C>0$ with
\begin{align*}
	\frac{1}{N}\sum_{j=1}^N \left|\Delta SAL (R_{1,j},R_{2,j},p_1,p_2)\right| \leq C \left|\Delta p\right|.
\end{align*}
While maximal changes in SAL scores represent worst case scenarios, the mean value of score changes presents a starting point to a distributional analysis of parameter sensitivity. We denote SAL as \textit{maximum-unstable} or \textit{mean-unstable} with respect to a parameter of an OIA, if no bounding constants exist, i.e. small changes in the parameter value can lead to large changes, or even unbounded responses, in the resulting SAL scores.\\

\begin{figure*}
\center
\includegraphics[width=16cm]{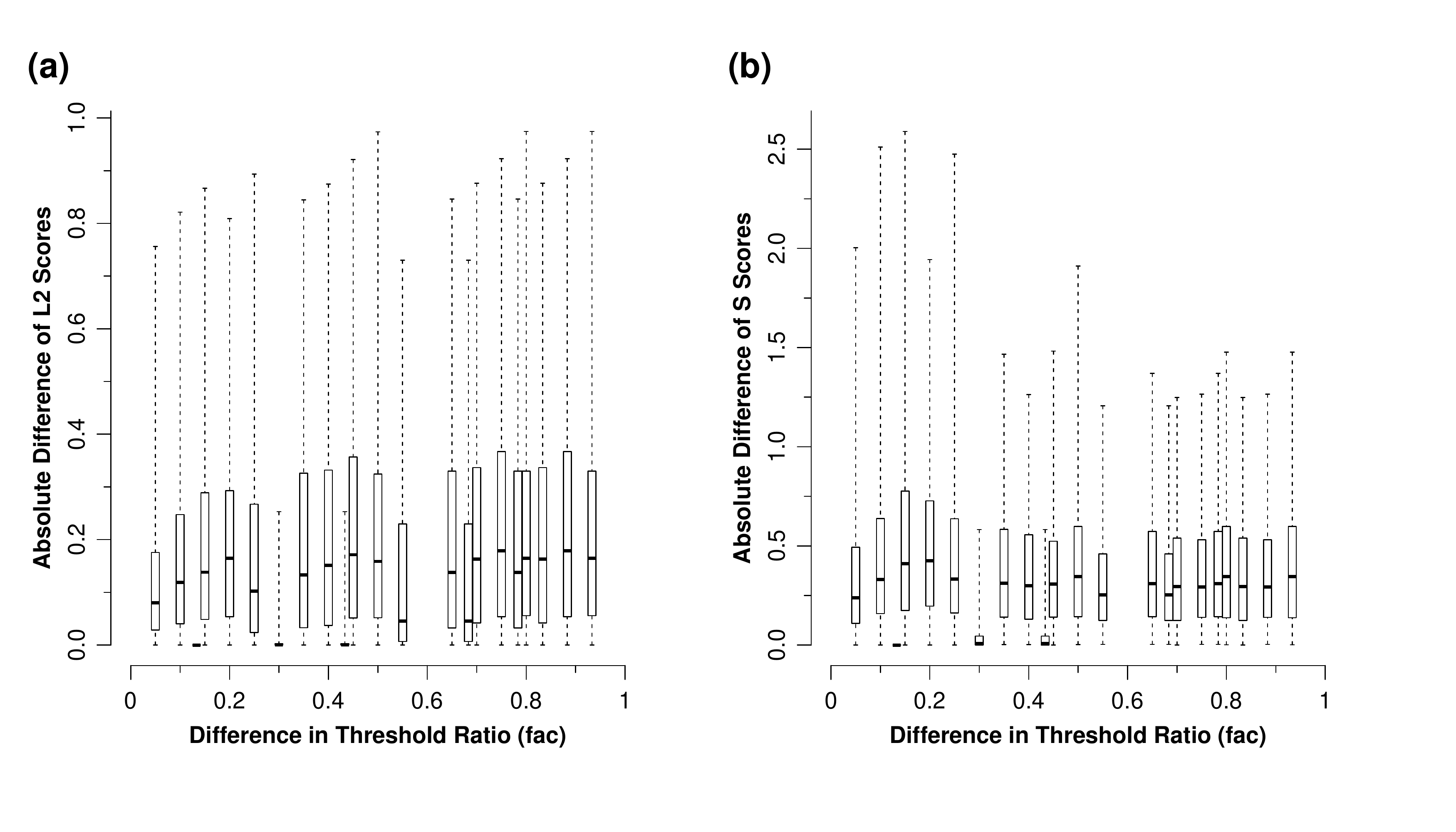}
\caption{
Differences in
(a) $L2$ and (b) $S$ scores for IR6.2 with respect to changes in the  threshold ratio (\textit{fac}) of the \textit{threshfac} algorithm. The box-whiskers represent the score differences over the 400 spatial fields. Solid boxes indicate the interquartile range, while the dashed lines reach out to the extremes.
}
\label{fig:par_vs_score_threshfac}
\end{figure*}

\begin{figure*}
\center
\includegraphics[width=16cm]{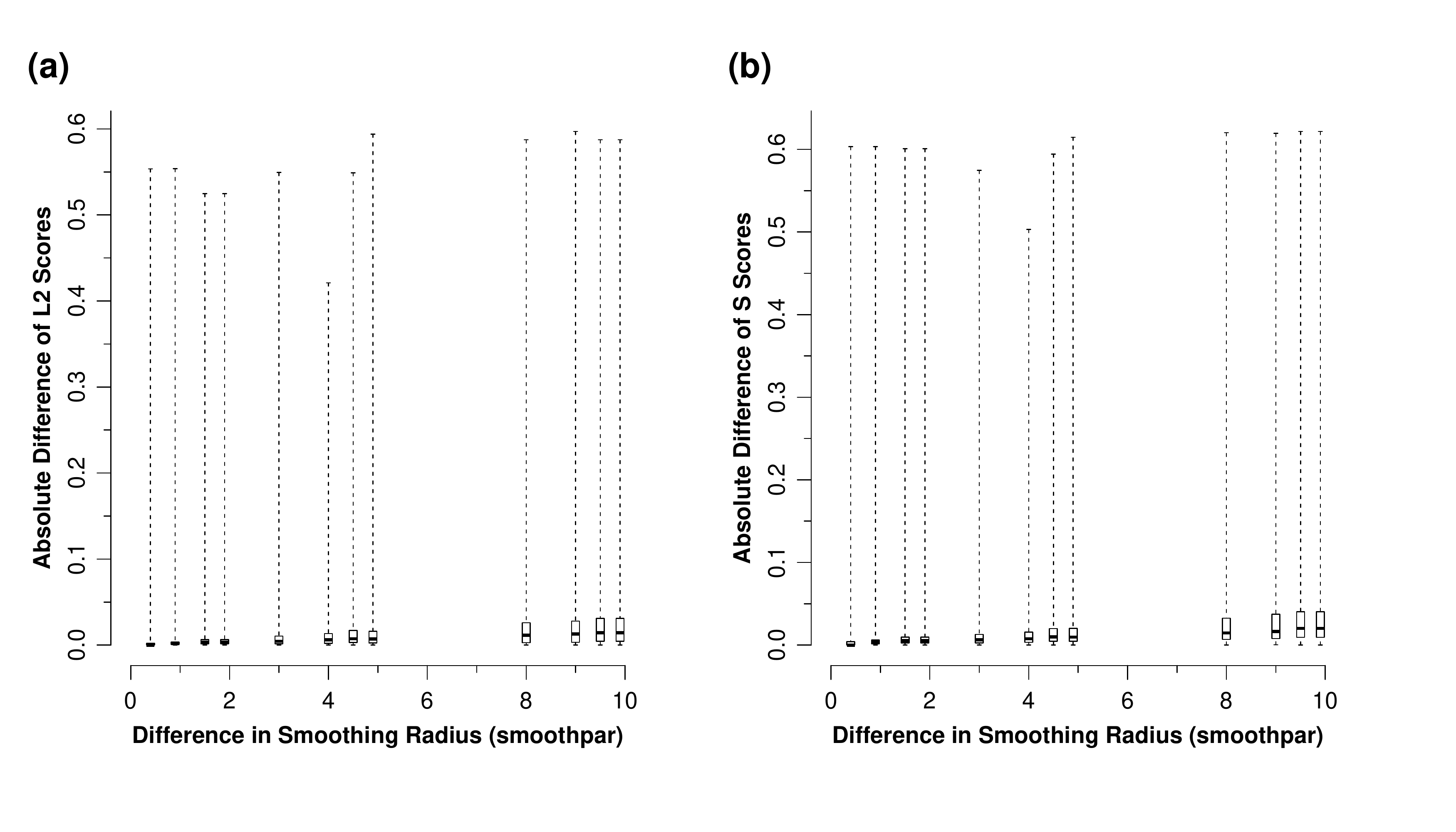}
\caption{
Same as Fig. \ref{fig:par_vs_score_threshfac} but with respect to changes in the  smoothing radius (\textit{smoothpar}) of the \textit{convthresh} algorithm. 
}
\label{fig:par_vs_score_convthresh}
\end{figure*}

\begin{figure*}
\center
\includegraphics[width=16cm]{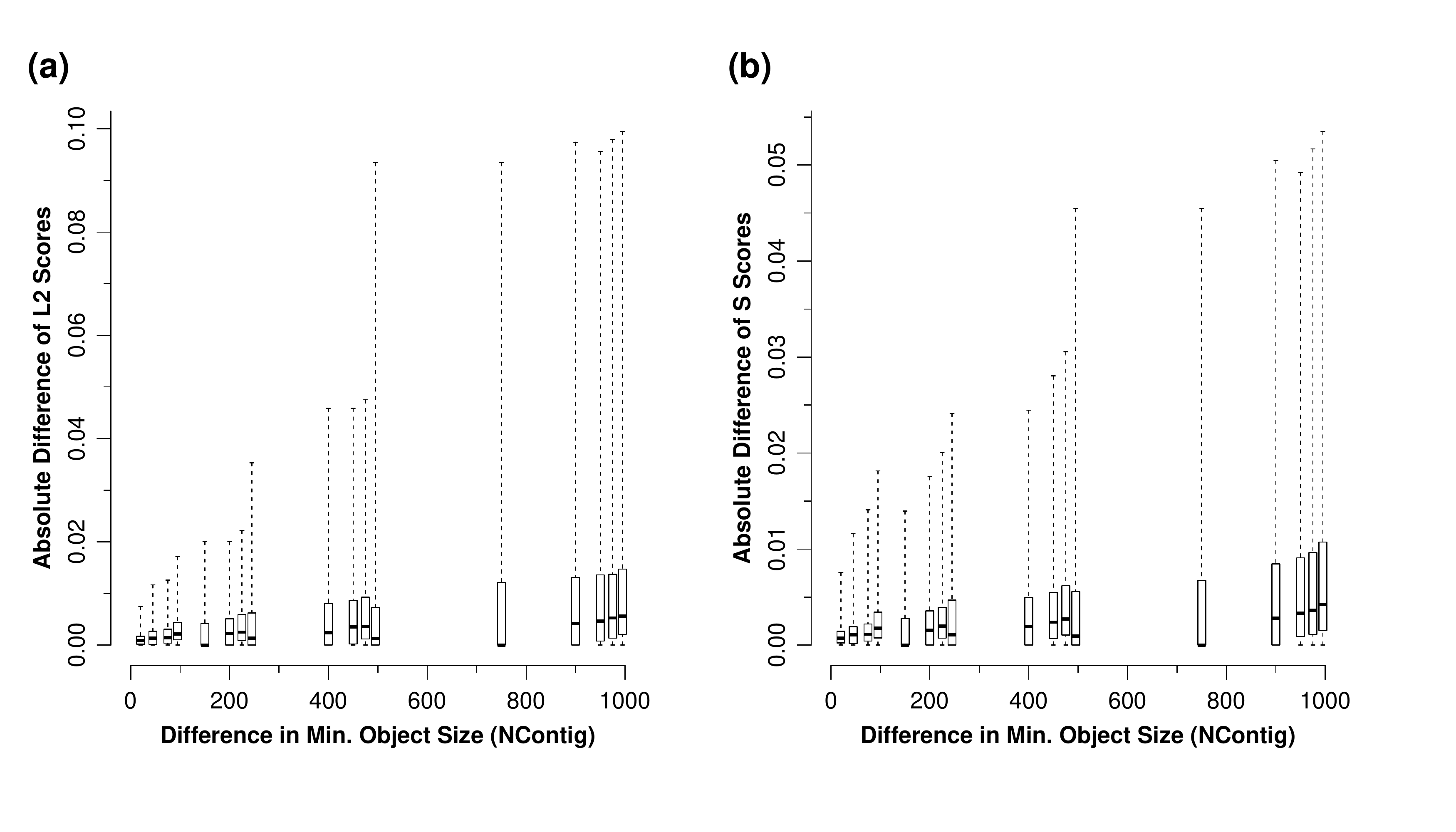}
\caption{
Same as Fig. \ref{fig:par_vs_score_threshfac} but with respect to changes in the  minimum object size \textit{NContig}) of the \textit{threshsizer} algorithm.}
\label{fig:par_vs_score_threshsizer}
\end{figure*}

Figs.~\ref{fig:par_vs_score_threshfac}-\ref{fig:par_vs_score_threshsizer} show the responses of $L2$ and $S$ scores to parameter changes of the OIA for IR6.2. For \textit{stable} parameters we expect a linear decrease in (maximal and mean) absolute score differences for decreasing differences in parameter value.

Fig.~\ref{fig:par_vs_score_threshfac} shows that the threshold ratio, which varies for the \textit{threshfac} algorithm, clearly induces \textit{maximum-unstable} and \textit{mean-unstable} SAL scores. The maximum of absolute differences in $L2$ score is as high as $1.0$. As discussed for the conceptual example this corresponds to the difference between the best and the worst score possible. This holds true for the $S$ score as well, with a maximum difference of $2.5$. Naturally, the effect on the mean value is smaller but still very significant at about $0.2$ for the $L2$ score and $0.4$ for the $S$ score.

The \textit{convthresh} algorithm with a varying smoothing radius and a fixed threshold ratio is studied in Fig.~\ref{fig:par_vs_score_convthresh}. Here, the results are more complex: while the maxima of absolute score differences indicate a \textit{maximum-unstable} behavior, the mean values indicate \textit{mean-stable} SAL scores. Largest $L2$ and $S$ score differences are of about $0.6$. These are smaller than in Fig.~\ref{fig:par_vs_score_threshfac} by a factor of $1.7$ for the $L2$ score and $4$ for the $S$ score. The differences in the mean values are smaller by a factor of $20$ for both scores. Conclusively, the worst cases for varying smoothing radii are less severe and occur less frequently than for the threshold ratio of the \textit{threshfac} algorithm. Note, however, that a different behavior might be obtained with another data set. Whether or not SAL can be regarded as stable with respect to changes in the smoothing radius depends on two aspects: first, the data, and second, the question one wishes to address with the SAL verification. If latter includes the interpretation of quantiles other than the median (e.g., the interquartile range), a more elaborate statistical analysis is necessary.

Fig.~\ref{fig:par_vs_score_threshsizer} shows the same stability analysis for the \textit{threshsizer} algorithm, where the \textit{NContig} parameter (i.e.\ the minimal size of objects) varies. Here, both maxima and mean values exhibit a \textit{stable} behavior of both scores. The differences are much smaller than for the previous cases. Interestingly, even large changes in the parameter values lead to small changes in $S$ and $L2$ scores. 

In summary, we have an \textit{unstable} behavior of SAL with respect to the threshold parameter and a \textit{stable} behavior with respect to the minimum object size. Varying smoothing radii seem to induce \textit{mean-stable} behavior, but may result in \textit{unstable} SAL scores for some cases, which calls for a closer look at the distributional properties of score changes.\\

In order to assess the sensitivity of the SAL scores to changes in the OIA parameters on a distributional level, we study the null hypothesis of equal distributions of the $S$ and $L2$ scores. The following OIA parameters are considered: threshold ratio $f \in \{ 1/15, 0.2, 0.5, 0.75, 0.85, 0.9, 0.95, 1 \}$, smoothing radius $smoothpar \in \{ 0, 1, 2, 5, 10 \}$ and minimal object size $NContig \in \{ 5, 25, 50, 100, 250, 500, 1000 \}$. The null hypothesis is tested using three complementary hypothesis tests, notably the
	Kolmogorov-Smirnov \citep[K,][]{kolmogorov1933,smirnov1939},
	median (M), and
	quantile (Q) test, where the test statistic is the interquartile distance (i.e.\ the distance between the $25\%$ and $75\%$ quantiles).
The M and Q tests are permutation tests \citep{good2000permutation} with $10,000$ iterations and are particularly interesting for the study of SAL, since both median and interquartile distance are important quantities for the interpretation of SAL scores \citep{wernli2008sal}. 

For large variations of the threshold ratio in the \textit{threshfac} algorithm one potentially looks at different physical situations. Thus,  significant differences in the distributions of the $S$ and $L2$ scores are expected for large differences $\Delta f > 0.1$. Table~\ref{tab:hypo-rad062-threshfac} summarizes the results of all hypothesis tests and all combinations of threshold ratios for $400$ fields of IR6.2 data. The upper right triangular shows results for the comparison of two $S$ score distributions (italic font), while the lower left triangular shows the results for $L2$ (bold font), e.g. for the two $S$ distributions derived with $f=0.75$ and $f=0.85$ only the quantile test (Q) indicates significant differences, while the $L2$ distributions with the same parameter settings differ in all three test statistics (KMQ). Table~\ref{tab:hypo-rad062-threshfac} confirms our expectations for IR6.2 for all but the largest threshold ratios of the \textit{threshfac} algorithm. These results are consistent with the stability analysis for the threshold ratio in the previous section. The Q test is tailored to detect changes in spread and therefore well suited for a two-sided score. Accordingly, only the Q test is able to distinguish between $S$ distributions for threshold ratios $0.75$ and $0.85$ (Table~\ref{tab:hypo-rad062-threshfac}). Note that changes in $S$ scores due to object decomposition are symmetric, since the decomposition can happen in both observations and/or forecasts. 

Both the \textit{convthresh} and \textit{threshsizer} algorithms show significant differences only in the distribution of the $L2$ but not the $S$ scores (Tables \ref{tab:hypo-rad062-convthresh} and \ref{tab:hypo-rad062-threshsizer}). The reason for this is twofold: first, $S$ is a two-sided score and changes in the score may cancel out in accumulated statistics like mean or median. Therefore, the M test is less powerful to detect changes in the $S$ score distribution. Second, while -- in principle -- the K test is able to detect changes in spread, it has issues when only the outer tails of the distributions are affected \citep[e.g.][]{mason1983}. 
In summary we can identify symmetric changes in distributions for the $S$ score only if they affect the interquartile distance. By definition, the interquartile distance is largely unaffected by small variation of individual values. Hence, changes in the $S$ score are expected to be small for the less critical parameters, which is consistent with the results observed in Tables \ref{tab:hypo-rad062-convthresh} and \ref{tab:hypo-rad062-threshsizer}.

The distributional analysis has confirmed the unstable behavior of the SAL scores with respect to the threshold ratio in the \textit{threshfac} algorithm. Whether or not smoothing radius and minimum object size can be considered as uncritical parameters depends on the interpretation of the SAL scores: if one is interested in the statistical quantities mean, median and interquartile distance both can be considered as giving fairly stable SAL scores for IR6.2. However, for  a varying smoothing radius this highly depends on the data. Table~\ref{tab:hypo-clct-convthresh} shows the results for TClC, where both median and interquartile distance change significantly for many parameter pairings.\\

\begin{table}[p]
\center
\begin{tabular}[ht]{ccccccccc} 
		\toprule[2pt]
		$f$    & $1/15$ & $0.20$& $0.50$& $0.75$& $0.85$& $0.90$& $0.95$& $1$ \\
		\midrule[1pt]
		$1/15$ &     & \cb -   & \cb Q   & \cb KMQ & \cb KMQ & \cb KMQ & \cb KMQ & \cb KMQ \\
		$0.20$ & \cy -   &     & \cb Q   & \cb KMQ & \cb KMQ & \cb KMQ & \cb KMQ & \cb KMQ \\ 
		$0.50$ & \cy KMQ & \cy KMQ &     & \cb KMQ & \cb KMQ & \cb KMQ & \cb KMQ & \cb KMQ \\
		$0.75$ & \cy KMQ & \cy KMQ & \cy KMQ &     & \cb Q   & \cb -   & \cb -   & \cb -   \\ 
		$0.85$ & \cy KMQ & \cy KMQ & \cy KMQ & \cy KMQ &     & \cb -   & \cb -   & \cb -   \\
		$0.90$ & \cy KMQ & \cy KMQ & \cy KMQ & \cy KMQ & \cy -   &     & \cb -   & \cb -   \\
		$0.95$ & \cy KMQ & \cy KMQ & \cy KMQ & \cy KMQ & \cy K   & \cy -   &     & \cb -   \\
		$1$    & \cy KMQ & \cy KMQ & \cy KMQ & \cy KMQ & \cy K   & \cy -   & \cy -   &     \\
 		\bottomrule[2pt]
\end{tabular}
\caption[Significant differences in score distributions for different values of the threshold ratio $f$ in the {\sl threshfac} algorithm for $400$ fields of IR6.2. The capital letters denote hypothesis tests (defined in Section~\ref{sec:analysis-of-parameter-sensitivity}) detecting differences at a $5\%$ level of significance. The upper right triangular shows results for the comparison of two $S$ score distributions (italic font), while the lower left triangular shows the results for $L2$ (bold font).]
{
Significant differences in score distributions for different values of the threshold ratio $f$ in the {\sl threshfac} algorithm for $400$ fields of IR6.2. The capital letters denote hypothesis tests (defined in Section~\ref{sec:analysis-of-parameter-sensitivity}) detecting differences at a $5\%$ level of significance. The upper right triangular shows results for the comparison of two $S$ score distributions (italic font), while the lower left triangular shows the results for $L2$ (bold font).}
\label{tab:hypo-rad062-threshfac}
\end{table}

\begin{table}[p]
\center
\begin{tabular}[ht]{cccccc} 
		\toprule[2pt]
		$smoothpar$ & $0$ &  $1$  &  $2$  &  $5$ & $10$  \\
		\midrule[1pt]
		$0$   &     & \cb -   & \cb -   & \cb -  & \cb -   \\
		$1$   & \cy -   &     & \cb -   & \cb -  & \cb -   \\
		$2$   & \cy K   & \cy -   &     & \cb -  & \cb -   \\ 
		$5$   & \cy K   & \cy K   & \cy K   &    & \cb -   \\
		$10$  & \cy KM  & \cy KM  & \cy KM  & \cy K  &     \\	
 		\bottomrule[2pt]
\end{tabular}
\caption[Significant differences in score distributions for different values of the smoothing radius $smoothpar$ in the {\sl convthresh} algorithm for $400$ fields of IR6.2. The capital letters denote hypothesis tests (defined in Section~\ref{sec:analysis-of-parameter-sensitivity}) detecting differences at a $5\%$ level of significance. The upper right triangular shows results for the comparison of two $S$ score distributions (italic font), while the lower left triangular shows the results for $L2$ (bold font).
]{
Significant differences in score distributions for different values of the smoothing radius $smoothpar$ in the {\sl convthresh} algorithm for $400$ fields of IR6.2. The capital letters denote hypothesis tests (defined in Section~\ref{sec:analysis-of-parameter-sensitivity}) detecting differences at a $5\%$ level of significance. The upper right triangular shows results for the comparison of two $S$ score distributions (italic font), while the lower left triangular shows the results for $L2$ (bold font).}
\label{tab:hypo-rad062-convthresh}
\end{table}

\begin{table}[p]
\center
\begin{tabular}[ht]{cccccccc} 
		\toprule[2pt]
		$NContig$ & $5$  & $25$ & $50$ & $100$& $250$& $500$& $1000$ \\
		\midrule[1pt]
		$5$     &    & \cb -  & \cb -  & \cb -  & \cb -  & \cb -  & \cb - \\
		$25$    & \cy -  &    & \cb -  & \cb -  & \cb -  & \cb -  & \cb - \\
		$50$    & \cy K  & \cy -  &    & \cb -  & \cb -  & \cb -  & \cb - \\
		$100$   & \cy K  & \cy K  & \cy K  &    & \cb -  & \cb -  & \cb - \\
		$250$   & \cy K  & \cy K  & \cy K  & \cy K  &    & \cb -  & \cb - \\
		$500$   & \cy KM & \cy KM & \cy K  & \cy K  & \cy K  &    & \cb - \\
		$1000$  & \cy KM & \cy KM & \cy KM & \cy KM & \cy KM & \cy K  &   \\
 		\bottomrule[2pt]
\end{tabular}
\caption[Significant differences in score distributions for different values of the minimum object size $NContig$ in the {\sl threshsizer} algorithm for $400$ fields of IR6.2. The capital letters denote hypothesis tests (defined in Section~\ref{sec:analysis-of-parameter-sensitivity}) detecting differences at a $5\%$ level of significance. The upper right triangular shows results for the comparison of two $S$ score distributions (italic font), while the lower left triangular shows the results for $L2$ (bold font). 
]{
Significant differences in score distributions for different values of the minimum object size $NContig$ in the {\sl threshsizer} algorithm for $400$ fields of IR6.2. The capital letters denote hypothesis tests (defined in Section~\ref{sec:analysis-of-parameter-sensitivity}) detecting differences at a $5\%$ level of significance. The upper right triangular shows results for the comparison of two $S$ score distributions (italic font), while the lower left triangular shows the results for $L2$ (bold font).}
\label{tab:hypo-rad062-threshsizer}
\end{table}

\begin{table}[p]
\center
\begin{tabular}[ht]{cccccc} 
		\toprule[2pt]
		$smoothpar$ & $0$ &  $1$  &  $2$  &  $5$ & $10$  \\
		\midrule[1pt]
		$0$   &     & \cb -   & \cb -   & \cb -  & \cb -     \\
		$1$   & \cy -   &     & \cb -   & \cb -  & \cb -     \\
		$2$   & \cy KM  & \cy -   &     & \cb -  & \cb -     \\ 
		$5$   & \cy KM  & \cy KM  & \cy KM  &    & \cb -     \\
		$10$  & \cy KMQ & \cy KMQ & \cy KMQ & \cy KM &       \\
		\bottomrule[2pt]
\end{tabular}
\caption[Significant differences in score distributions for different values of the smoothing radius $smoothpar$ in the {\sl convthresh} algorithm for $400$ fields of TClC. The capital letters denote hypothesis tests (defined in Section~\ref{sec:analysis-of-parameter-sensitivity}) detecting differences at a $5\%$ level of significance. The upper right triangular shows results for the comparison of two $S$ score distributions (italic font), while the lower left triangular shows the results for $L2$ (bold font).]
{
Significant differences in score distributions for different values of the smoothing radius $smoothpar$ in the {\sl convthresh} algorithm for $400$ fields of TClC. The capital letters denote hypothesis tests (defined in Section~\ref{sec:analysis-of-parameter-sensitivity}) detecting differences at a $5\%$ level of significance. The upper right triangular shows results for the comparison of two $S$ score distributions (italic font), while the lower left triangular shows the results for $L2$ (bold font).}
\label{tab:hypo-clct-convthresh}
\end{table}

\begin{figure*}
\center
\includegraphics[width=16cm]{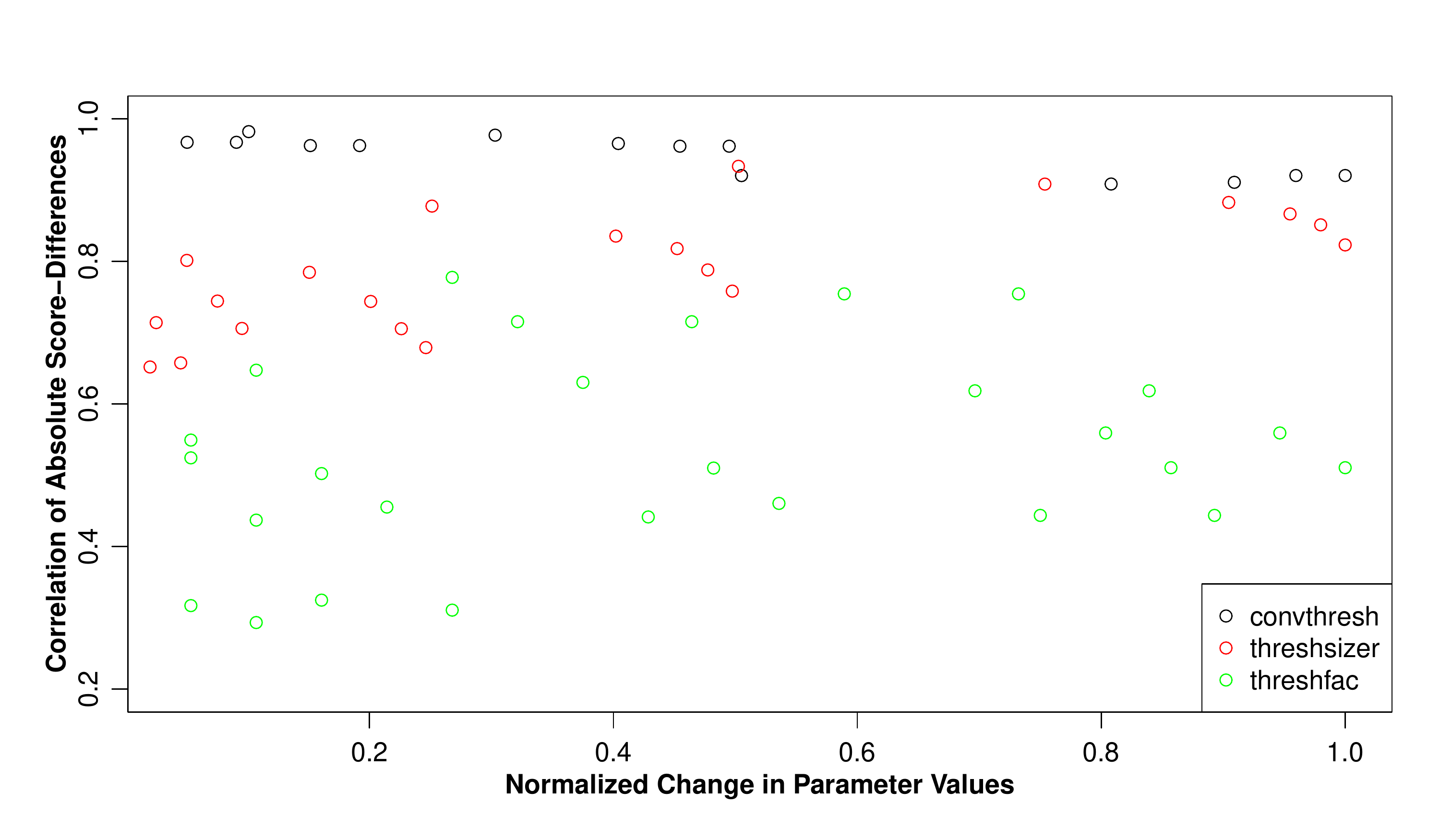}
\caption{
The Pearson correlation coefficients between absolute changes in $S$ and $L2$ scores for IR6.2 are plotted against the normalized difference in parameter values. For the \textit{convthresh} algorithm we only vary the smoothing radius (\textit{smoothpar}), for the \textit{threshsizer} algorithm only the minimal object size (\textit{NContig}) and for the \textit{threshfac} algorithm the threshold ratio (\textit{fac}).
}
\label{fig:corr_combined_rad062}
\end{figure*}

It is often stated that the components of SAL are independent \citep[e.g.][]{fruh2007verification,zimmer2008feature}. However, this is not true in the sense of the mathematical definition of statistical independence. Quite contrarily, Fig.~\ref{fig:corr_combined_rad062} shows significant Pearson correlation coefficients between the absolute $L2$ and $S$ differences of each of the 400 fields for all algorithms and standardized parameter changes (i.e. the difference of two parameter values is divided by the maximum difference we investigated for this parameter). The correlation coefficients between the absolute $L2$ and  $S$ differences are even close to one for the \textit{convthresh} algorithm, where most significant changes are due to object decomposition (see Section \ref{sec:oia}). The decomposition of a large object into two, or the emergence of more small objects impacts the structure ($S$) and the spread of objects ($L2$), simultaneously. 

The correlation coefficients for the \textit{threshsizer} algorithm are slightly lower with values between $0.7$ and $0.9$. Here, an increasing \textit{Ncontig} parameter implies that objects of increasing size are omitted. Since only the small-scale objects are affected, the structure is shifted towards larger objects with more mass. At the same time the spread is reduced. For \textit{threshfac} we observe a large spread in the correlation coefficients varying between $0.4$ and $0.8$. This underlines the unstable behavior with respect to a varying threshold level, since many different effects can occur: small changes can lead to object decomposition, while large changes can cause objects of arbitrary size to vanish. Latter can lead to vastly different $S$ scores, but nearly unchanging $L2$ scores, which in turn leads to lower correlation coefficients. This behavior can be observed for the exemplary case in Figure \ref{fig:case_study_rad062_threshfac}.

\section{Indicators for Parameter Sensitivity and Observational Uncertainty}
\label{sec:indicators}

Of the three investigated OIA parameters the threshold ratio is most important one for two reasons. First, all three OIA depend on the threshold level. Second, the SAL scores show the largest sensitivity to changes in the threshold ratio. Therefore, we concentrate solely on the \textit{threshfac} algorithm in this section. Since the calculation of SAL scores for a multitude of different threshold ratios rapidly becomes computationally expensive, it is useful for practical applications to find a quantity that indicates whether or not a given set of data exhibits a high sensitivity towards small variations in threshold ratios, i.e. whether SAL is \textit{mean-stable} with respect to threshold ratio. To derive such a sensitivity indicator we concentrate on changes in the $L2$ score, which is mathematically more accessible. Section \ref{sec:analysis-of-parameter-sensitivity} shows that the results hold true also for the $S$ score. 

\begin{figure*}
\center
\includegraphics[width=\textwidth]{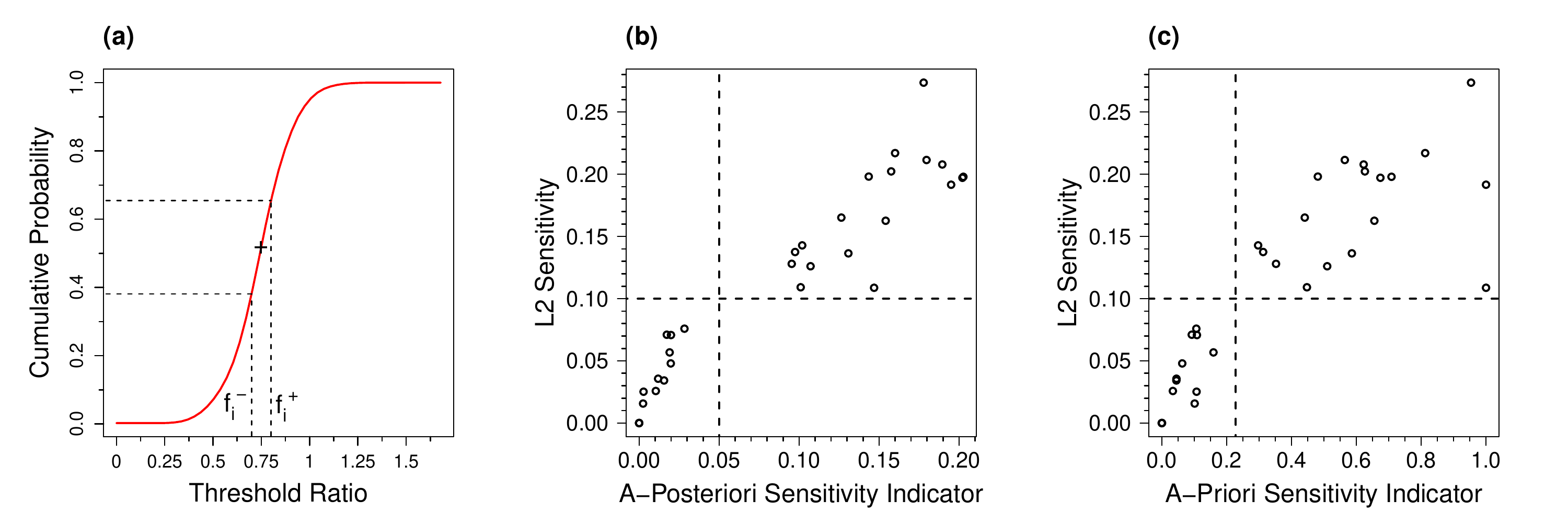}
\caption{
Sensitivity Indicators: (a) Univariate ECDF of 400 fields of IR6.2. For a given threshold ratio of $0.75$ (black cross), varying threshold ratios ($f_i^-$, $f_i^+$) and their resulting ECDF-values are marked with dashed lines. Panel (b) shows $L2$ sensitivity against the a-posteriori sensitivity indicator $SI$, and (c) against the a-priori indicator $\widetilde{SI}$.
}
\label{fig:indicators}
\end{figure*}

Since we are interested in the response to small parameter changes, we vary each of the eight threshold ratios $f \in \{ 1/15, 0.2, 0.5, 0.75, 0.85, 0.9, 0.95, 1 \}$ additionally by $\pm 0.05$, and calculate $L2$ scores for all 24 resulting parameter values. Lets denote the original threshold level by $f_i$, the perturbed level by $f_i^-$ and $f_i^+$, and the resulting $L2$ scores by $L2_i^-$, $L2_i$ and $L2_i^+$, for $i \in \{1, ..., 8\}$. The $L2$ sensitivity at threshold level $f_i$ for a single field-to-field comparison ($R^{fcst}$ vs. $R^{obs}$) is given by the diameter (i.e. the maximum pairwise distance) of the set $\{ L2_i^-(R^{fcst},R^{obs}), L2_i(R^{fcst},R^{obs}), L2_i^+(R^{fcst},R^{obs}) \}$. Taking the mean value over $N=400$ field-to-field comparisons then yields the $L2$ sensitivity ($\delta_{L2}$) at threshold level $f_i$ for the complete set of data:
\begin{align*}
	\delta_{L2}(f_i) = \frac{1}{N}\sum_{l=1}^N  \text{diam}\left( \left\{ L2_i^-(R_l^{fcst},R_l^{obs}), L2_i(R_l^{fcst},R_l^{obs}), L2_i^+(R_l^{fcst},R_l^{obs}) \right\} \right).
\end{align*}
Largest sensitivity is expected in cases where a slight increase in the threshold ratio causes a large number of grid points to fall below the threshold. We are therefore interested in the ratio of grid points that vanish for a given field due to an increase in threshold ratio. This quantity can be approximated via the univariate empirical cumulative distribution function (ECDF) of the total set of spatial fields (see Fig.~\ref{fig:indicators} a), which describes the probability that IR6.2 is below a threshold. The ECDF of observational data is defined as
\begin{align*}
	\text{ecdf}_1(t) = \frac{1}{N\ \#(\mathcal{D})}\sum_{l=1}^N \#\left(\left\{x\in\mathcal{D}\ \big|\ R^{obs}_l(x)\leq t\right\}\right),
\end{align*}
where $\#(\cdot)$ denotes the number of elements in a set. The ECDF of the set of forecasts fields is denoted by ecdf$_2$ and defined analogously. The ECDF-ratios, which are functions of the parameter value $f_i$, are given by
\begin{align*}
	p^k_i &= \frac{\text{ecdf}_k\left(f_i^+\right) - \text{ecdf}_k\left(f_i^-\right)}{1 - \text{ecdf}_k\left(f_i^-\right)}, \quad k \in \{1,2\},\ i \in \{1, ..., 8\}.
\end{align*}
The lowest threshold $f^{-}_i$ is used in the denominator to ensure that the ratio has an upper bound equal to one. This is necessary to allow the interpretation of $p^k_i$ as a first order approximation of a ``decomposition-probability'', i.e. the probability for the event that a large object decomposes into two or more smaller objects. This interpretation is intuitive as seen for the following two extreme cases. First, if no point vanishes when raising the threshold, the probability that a large object decomposes is zero and $p^k_i=0$. Second, if all points vanish, the probability that a large object vanishes is one and $p^k_i=1$. Therefore, $p^k_i$ approximates the decomposition probability while ignoring any effects of spatial correlation. This helps us to estimate the sensitivity of $L2$, since we can now quantify the probability that object decomposition occurs and hence sensitivity is high. It may seem overly simplified to use a single ECDF for a whole set of $400$ spatial fields instead of $400$ independent ECDFs. However, we are not interested in single worst case scenarios, but in an indicator for \textit{mean-stability}, which is a statistical quantity depending on the average score deviations in the data set. This justifies the above approach as a reasonable and computational effective first guess\footnote{To be on the safe side, we have also calculated indicators analogous to $SI$ and $\widetilde{SI}$ defined in \eqref{eq:SI} and \eqref{eq:tSI} but based on single-field ECDFs. However, no significant improvements could be observed.}.

We further need to quantify the effect that a process of object decomposition has on $L2$. Recall that the $L2$ score at threshold level $f_i$ is defined as
\begin{align*}
	L2_i &= 2\frac{| r_1(f_i) - r_2(f_i) |}{d}, \quad i \in \{1, ..., 8\},
\end{align*}
where $r_1$ and $r_2$ describe the scattering of objects (see Section~\ref{sec:definition-of-sal}). $L2_i$ as well as $r_1(f_i)$ and $r_2(f_i)$ are statistical estimators. The variance -- or standard deviation -- of a statistical estimator is closely related to its robustness. We therefore use the empirical standard deviation $\sigma$ over the whole set of spatial fields of the three following quantities as a measure of the effect an object decomposition would have on the $L2$ score. The effect is quantified as
\begin{itemize}
	\item $2r_1/d$, if object decomposition occurs only in the first spatial field.
	\item $2r_2/d$, if object decomposition occurs only in the second spatial field.
	\item $L2_i$, if object decomposition occurs in both spatial fields simultaneously.
\end{itemize}
Note that all standard deviations are calculated based only on SAL values for the threshold $f_i$ and do not use any SAL values or calculations for the perturbed thresholds $f_i^-$ and $f_i^+$. 

Using the ECDF-ratios $p_i^k$ as first order approximation for the ``decomposition probabilities'' we can estimate the expected $L2$ sensitivity for threshold $f_i$ as follows
\begin{align}
\label{eq:SI}
	SI_i &= p^1_i (1-p^2_i)\  \sigma\left(\frac{2r_1(f_i)}{d}\right) + p^2_i (1-p^1_i)\ \sigma\left(\frac{2r_2(f_i)}{d}\right) + p^1_i p^2_i\ \sigma\left(L2_i\right),
\end{align}
for $i \in \{1, ..., 8\}$. Figure~\ref{fig:indicators}.b shows that $SI$ is indeed a good indicator for $L2$ sensitivity: if $SI<0.05$, the change of $L2$ is bounded by $0.1$. Note that once $L2$ scores have been calculated for a given threshold, $SI$ can be computed with very little computational cost, since all its components but the univariate ECDF-ratios have already been calculated for $L2$. For an a-priori indicator that uses only the ECDF information and no SAL values, we set $2r^1_i/d$, $2r^2_i/d$ and $L2_i$ equal one and obtain
\begin{flalign}
\label{eq:tSI}
	\widetilde{SI}_i &= p^1_i (1-p^2_i)\  + p^2_i (1-p^1_i)\ + p^1_i p^2_i,\quad i \in \{1, ..., 8\}.
\end{flalign}
Figure~\ref{fig:indicators}.c shows that the correlation between the $L2$ sensitivity and $\widetilde{SI}$ is still strong. This indicates that a large part of the stability issues is founded in the univariate field-distributions, i.e. the slopes of the univariate ECDFs, while spatial correlations only play a minor role.\\

There exists a close link between the sensitivity of SAL to varying threshold levels and the effects of observational uncertainties: if we look at the thresholded field, it makes no difference whether we raise the threshold level by a certain amount or lower the intensity of the field by the same constant amount at each grid point. The latter is equivalent to the effect of observational uncertainties with infinite spatial correlation length, i.e. an additive constant. Therefore, the results for the sensitivity of SAL to varying threshold levels carry over to its sensitivity to large scale uncertainties. 

Do the results also contain valuable information about the effect of small scale uncertainties? The previous section shows that $\widetilde{SI}$ is a good indicator for the $L2$ sensitivity with infinite spatial correlation length. However, $\widetilde{SI}$ employs only univariate ECDF information of the fields, i.e. it ignores any information regarding spatial correlations. Therefore, the correlation length of the (observational) uncertainties can only play a minor role for the sensitivity of SAL. This strongly suggests that the sensitivity of SAL to uncertainties of arbitrary correlation length is close to SAL's sensitivity to varying threshold levels. Consequently, $SI$ and $\widetilde{SI}$ are good indicators not only for the sensitivity of SAL towards varying threshold levels, but also for the sensitivity of SAL towards observational uncertainties.

\section{Discrimination power of SAL}
\label{sec:randomly-permuted-forecasts}
The sensitivity of the SAL parameters is closely linked to the ability of SAL to discriminate 'good' and 'bad' forecasts. We have shown that the SAL parameters are sensitive to changes in the OIA, and as argued in Section \ref{sec:indicators} a similar effect is expected for observational uncertainties. It is thus of interest to investigate the sensitivity of SAL scores towards artificial changes in the data itself. A somewhat savage way to produce an artificially 'bad' set of forecasts is to destroy the temporal collocation of the $400$ pairs of forecasted and observed fields by a random permutation of the fields in time, i.e. for each observation a random forecast is drawn (from the set of 400 available forecast fields). We then ask whether SAL is able to distinguish the quality of the original forecast and the randomly permuted forecast. This is achieved by testing the null hypothesis that the SAL scores over the 400 pairs of fields in the original and permuted data set follow the same distribution.

\begin{figure*}
\center
\includegraphics[width=16cm]{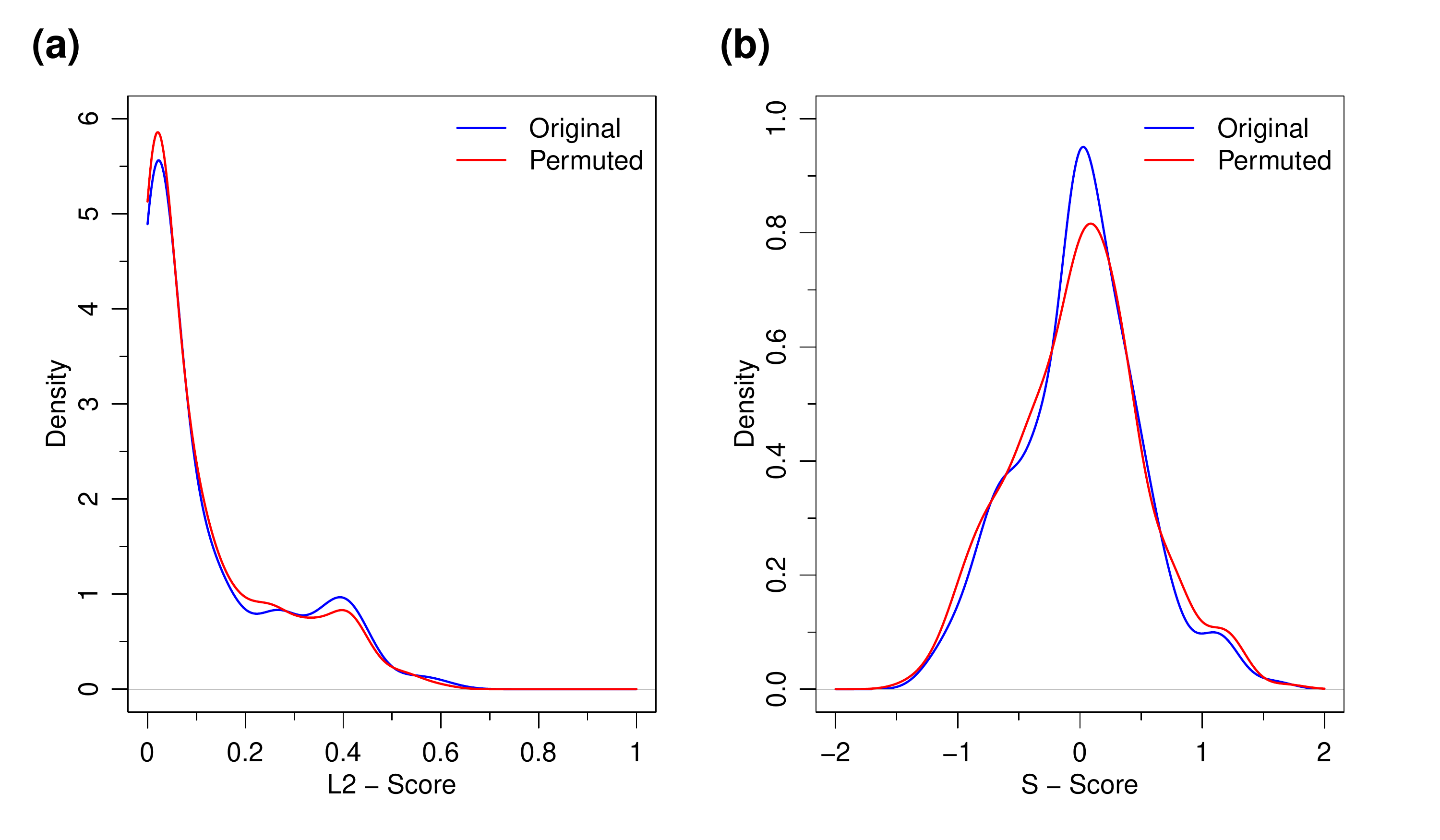}
\caption{
Density of (a) $L2$ and (b) $S$ score of original and permuted TClC with the default threshold ratio $f = 1/15$.
}
\label{fig:rand-perm-clct}
\end{figure*}

Table~\ref{tab:rand-perm} provides results of the different hypothesis tests on the SAL values using the \textit{threshfac} OIA with different threshold ratios for the two different variables TClC and IR6.2. We included TClC in this analysis to demonstrate that the results are strongly dependent on the type of data. For TClC and IR6.2 the default threshold ratio of $1/15$ exhibits almost no differences between both distributions (see Fig.~\ref{fig:rand-perm-clct}). Only the quantile permutation test is able to significantly detect differences in the $S$ score distribution of the IR6.2 data.

\begin{figure*}
\center
\includegraphics[width=16cm]{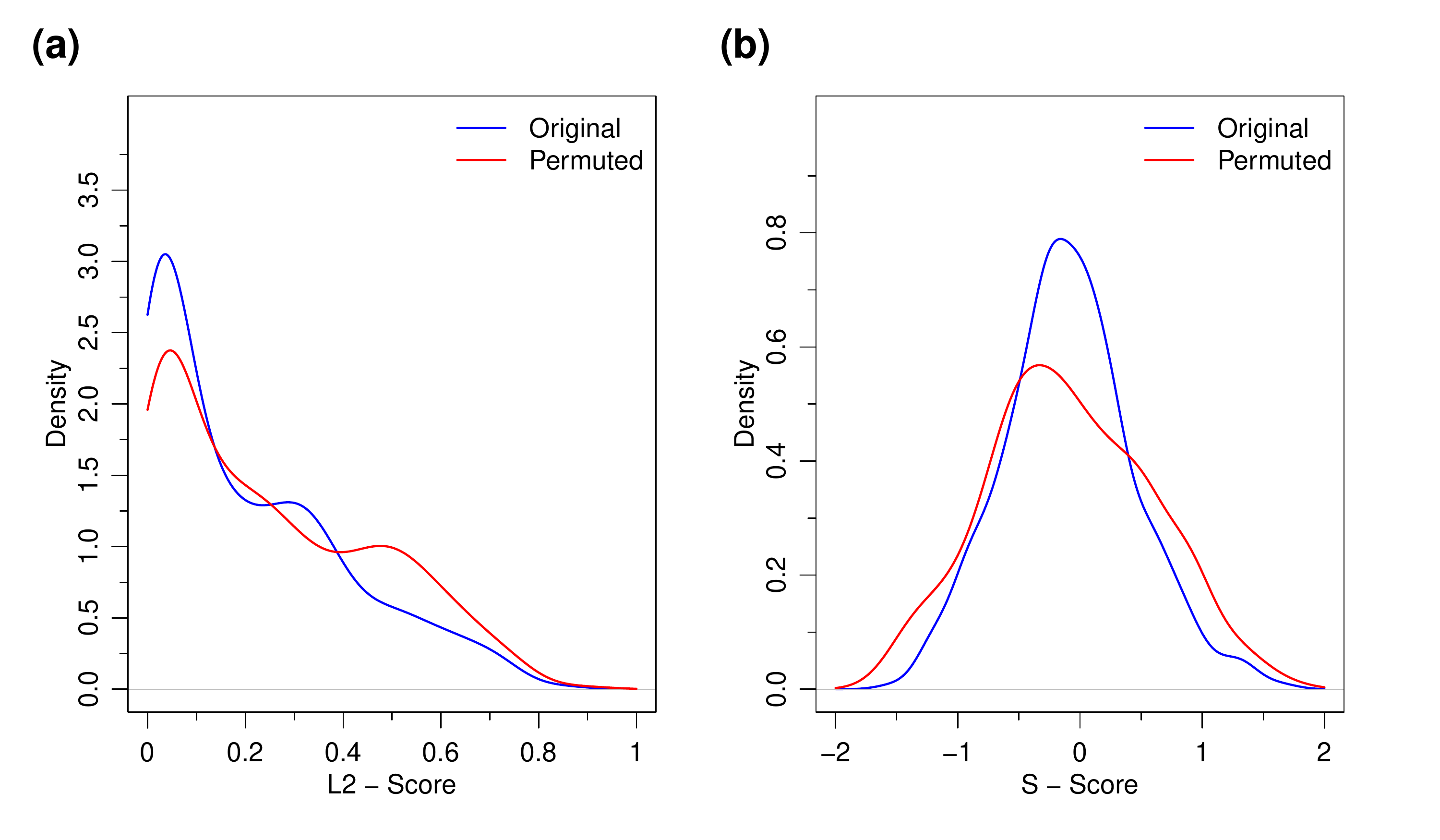}
\caption{
Density of (a) $L2$ and (b) $S$ score of original and permuted IR6.2 with threshold ratio $f = 0.85$.
}
\label{fig:rand-perm-rad062-S}
\end{figure*}

Fig.~\ref{fig:rand-perm-rad062-S} shows the density of the $L2$ and $S$ score distributions for IR6.2 for a higher threshold ratio of $0.85$. Although significant statistical differences can be observed for this threshold, the distributions do not look as different as one would expect for such a hyperbolic case. 
 
\begin{figure*}
\center
\includegraphics[width=8cm]{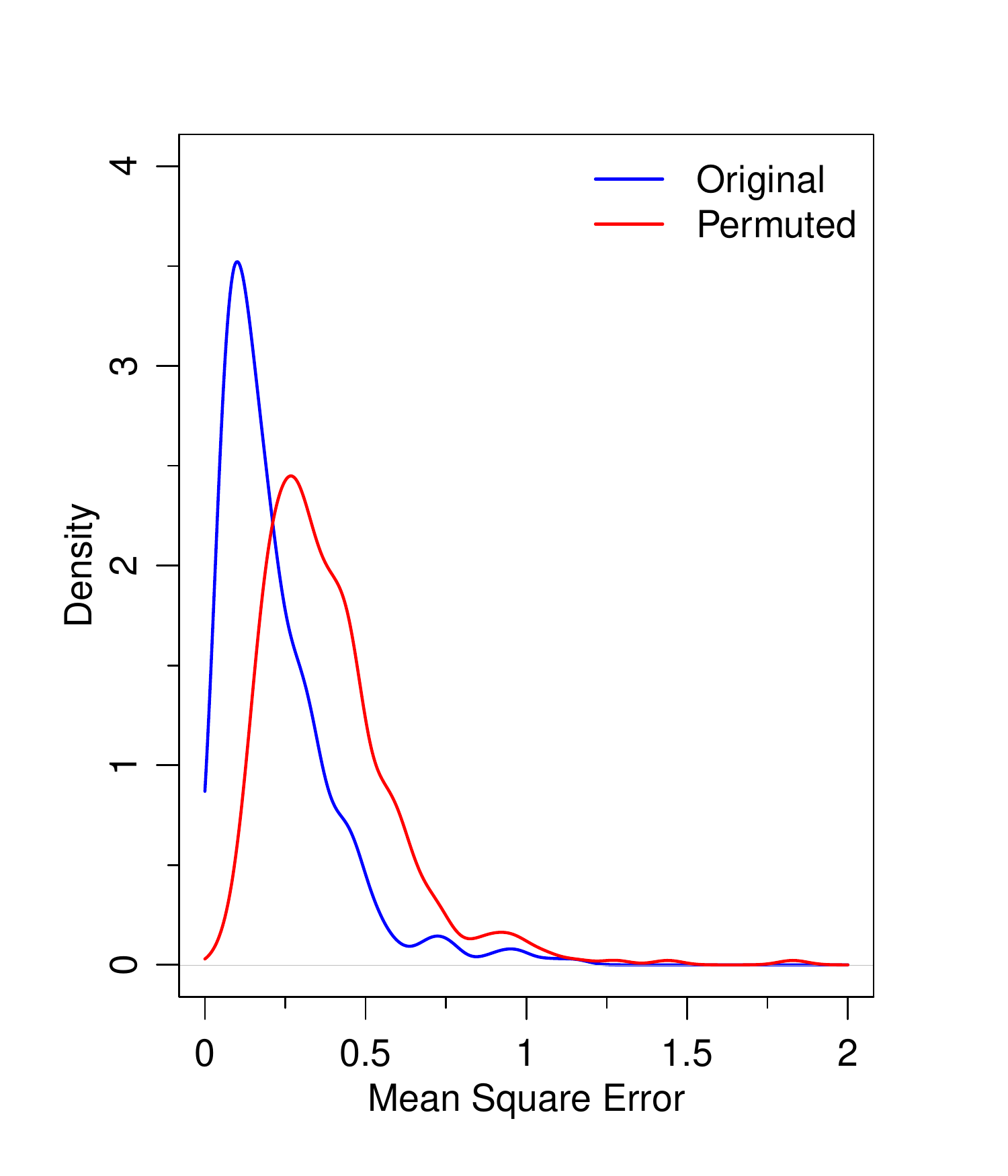}
\caption{Density of MSE for IR6.2 data of forecast vs. original and randomly permuted observations. This traditional score is able to distinguish both sets of data easily.}
\label{fig:rand-perm-mse}
\end{figure*}

The traditional mean square error (MSE) is able to clearly discriminate between the original and the permuted set of IR6.2 data, as shown in Fig.~\ref{fig:rand-perm-mse}. In the original data set the MSE is significantly smaller than in the permuted data set. 
 
For TClC and threshold ratios of $1/15$ or $0.2$ none of the two object dependent scores are able to discriminate between the original and the perturbed data set (Tab.~\ref{tab:rand-perm}). This changes for higher thresholds between $0.5$ and $0.95$, which suggests that the loss of discriminating power is due to very large objects at low thresholds. However, the situation is not as clear cut: the highest studied threshold ratio of $1$, which identified the smallest objects of all OIA settings, again yields indistinguishable score distributions. 

Conclusively, the ability of object dependent SAL scores to distinguish between the two data sets largely depends on the data, and cannot be guaranteed a-priori. It is important to note, that randomly permuting the observations is synonymous with a complete loss of temporal collocation between forecast and observation, which is expected to lead to significantly worse scores, as is the case for the MSE (Fig.~\ref{fig:rand-perm-mse}).

On its own, the inability to distinguish between the original and permuted data set is not necessarily a disaster. SAL investigates the statistics of the objects, and one might conclude that this is relatively homogeneous in time and thus insensitive to the loss of temporal collocation. However, in view of the large sensitivity the SAL scores show with respect to the choice of the OIA parameters, caution is advised when interpreting the results of SAL.

\begin{table}[p]
\center
\begin{tabular}[ht]{cccccc} 
		\toprule[2pt]
			$f$ & $S$(TClC) & $S$(IR6.2) & $L2$(TClC) & $L2$(IR6.2) \\
		\midrule[2pt]
			 $1/15$ & \cb -  & \cb Q  & \cy -   & \cy -\\
			 $0.20$ & \cb -  & \cb Q  & \cy -   & \cy -\\
			 $0.50$ & \cb -  & \cb Q  & \cy KQ  & \cy K\\
			 $0.75$ & \cb Q  & \cb KQ & \cy KM  & \cy KMQ\\
			 $0.85$ & \cb Q  & \cb KQ & \cy KMQ & \cy KMQ\\
			 $0.90$ & \cb KQ & \cb Q  & \cy KMQ & \cy KMQ\\
			 $0.95$ & \cb KQ & \cb -  & \cy KMQ & \cy KM\\
			 $1$    & \cb -  & \cb Q  & \cy -   & \cy KM\\
 		\bottomrule[2pt]
\end{tabular}
\caption{Significant differences between score distributions of $400$ fields derived from original observations vs. observations randomly permuted in time at various threshold ratios $f$ in the {\sl threshfac} algorithm. The capital letters denote hypothesis tests defined in Section~\ref{sec:analysis-of-parameter-sensitivity} that where able to detect differences with a $5\%$ level of significance. Results are shown for TClC and IR6.2 for the $S$ score (italic font) and $L2$ score (bold font).}
\label{tab:rand-perm}
\end{table}

\section{Conclusions}
\label{sec:conclusion}
The aim of this work is twofold: first, to study the applicability of SAL for cloud processes; second, to identify and understand the importance of OIA and their parameters on feature based verification methods and the link to observational uncertainties. Three different OIA have been used for the comparison of COSMO-DE SynSat data with SEVIRI satellite observations. In this process varying values have been used for the three parameters threshold ratio,  smoothing radius and minimal size of objects. On a conceptual level we have shown, that small changes in threshold levels or smoothing radii can potentially lead to very large score differences due to object decomposition, which is confirmed by two exemplary case studies of IR6.2 data.

To study SAL's parameter sensitivity on a distributional level, we denote SAL as \textit{stable} with respect to a parameter of an OIA if small changes in the value of this parameter can lead to large changes, or even unbounded responses, in the resulting SAL scores. SAL is \textit{unstable} with respect to threshold ratio and \textit{stable} with respect to minimal object size. With respect to varying smoothing radii SAL is \textit{mean-stable} but \textit{maximum-unstable}, i.e. there are rare worst case scenarios where large score deviations occur. In-depth statistical analysis using three different hypothesis tests confirms these results. For varying threshold ratios the observed large score deviations translate into significant changes in the distribution of $S$ and $L2$ scores. Consistent with the prior stability assessment the statistical implications for varying smoothing radii are much weaker.

The threshold ratio is of particular interest, not only because it is the most sensitive parameter, but because it links the field of parameter sensitivity to observational uncertainties: in cases where the intensity of a spatial field is close to the threshold level, changes in the threshold ratio (parameter sensitivity) lead to similar results as changes in the intensity of the data itself (observational uncertainties). An a-posteriori indicator ($SI\:$) for the stability of SAL to the threshold ratio parameter shows promising results to assess the sensitivity without the need of expensive computations of multiple threshold levels (Section~\ref{sec:indicators}). The a-priori indicator $\widetilde{SI}$ is based solely on univariate ECDF information of the spatial fields and can therefore be calculated with very little computational effort. Both quantities can also be employed to asses SAL's sensitivity to observational uncertainties (see Section~\ref{sec:indicators}).

Highly sensitive parameters are particularly problematic if the changes in scores due to varying parameters outweighs score deviations caused by actual differences in the data. Such a case is discussed in Section~\ref{sec:randomly-permuted-forecasts}, where $S$ and $L2$ scores where unable to reliably detect the complete loss of temporal allocation between forecast and observation. 

To summarize, the choice of OIA and its parameters has a significant effect on the resulting SAL scores. Therefore it is essential to explicitly state the algorithm and all parameter settings when using SAL for verification. The use of complementary hypothesis tests has shown that it is advisable to include statistical quantities beside median and interquartile range for the interpretation of SAL scores. The high sensitivity towards the threshold level implies a potentially high impact of observational uncertainties. This particularly true for SAL's original field of application, i.e. the verification of quantitative precipitation fields against radar observations. By defining sensitivity indicators ($SI\:$) and $\widetilde{SI}$, we were able to quantify the connection between parameter sensitivity and the effect of observational uncertainties. The fact, that small changes in parameter values have a larger impact than even drastic changes in data implies that SAL is not well equipped too verify this specific kind of data.

On a more technical level, object decomposition in conjunction with non-continuous operations, e.g. thresholding, during object identification has been established as the major cause for unstable behavior. Due to the similarity between threshold sensitivity and observational uncertainties, this study is a step towards the construction of feature based verification methods, that are robust with respect to observational uncertainties. The importance of the univariate ECDFs in the definition of ($SI\:$) and $\widetilde{SI}$ suggests, that the normalization of continuous data or the application of thresholds solely based on quantiles could significantly reduce the high sensitivity to parameters and uncertainties.

\acknowledgments
We gratefully acknowledge financial funding by the project High Definition Clouds and Precipitation for advancing Climate Prediction \hdcp funded by the German Ministry for Education and Research (BMBF) under grant FK 01LK1209B. The authors thank Sonja Reitter (Universit\"{a}t K\"{o}ln) and the DWD, who provided data from the COPS/GOP project, which was founded by the Deutsche Forschungsgemeinschaft under grant WU 356/4-2. We further appreciate the help of Jennifer Slobodda and Justus Franke (Institut f\"{u}r Weltraumwissenschaften, Freie Universit\"{a}t Berlin), who prepared SEVIRI data as part of the DFG-ICOS program.





 \bibliographystyle{ametsoc2014}
 \bibliography{mweniger}

\end{document}